\documentclass[fleqn,usenatbib]{mnras}

\usepackage{newtxtext,newtxmath}
\usepackage[T1]{fontenc}

\DeclareRobustCommand{\VAN}[3]{#2}
\let\VANthebibliography\thebibliography
\def\thebibliography{\DeclareRobustCommand{\VAN}[3]{##3}\VANthebibliography}

\usepackage{graphicx}	% Including figure files
\usepackage{amsmath}
\usepackage{amssymb}
\usepackage{newtxmath}
\usepackage{ragged2e}
\newcommand{\ropt}{$R_{\rm opt}^{\rm disc}~ $}
\newcommand{\rbreak}{$R_{\rm br} ~ $}
\newcommand{\rbreakk}{$R_{\rm br}$}

\newcommand{\reff}{$R_{\rm eff}~ $}
\newcommand{\rmin}{$R_{\rm min}~ $}
\newcommand{\rmax}{$R_{\rm max}~ $}

\title[Exploring the outskirts of the EAGLE disc galaxies]{Exploring the outskirts of the EAGLE disc galaxies}

\author[Varela-Lavin et al.]{
Silvio Varela-Lavin$^{1,2}$\thanks{E-mail: silvio.varela@userena.cl},
Patricia B. Tissera$^{3,4}$\thanks{Affiliated member of ARC Centre of Excellence for All Sky Astrophysics in 3 Dimensions (ASTRO 3D).},
Facundo A. G\'omez$^{1,2}$,
Lucas A. Bignone$^{5}$,\newauthor
Claudia del P. Lagos$^{6,7}$
\\
$^{1}$Departamento de F\'isica y Astronom\'ia, Universidad de La Serena, Av. Juan Cisternas 1200 Norte, La Serena, Chile.\\
$^{2}$ Instituto de Investigaci\'on Multidisciplinar en Ciencia y Tecnolog\'ia, Universidad de La Serena, Ra\'ul Bitr\'an 1305, La Serena, Chile.\\
$^{3}$Instituto de Astrof\'{i}sica, Pontificia Universidad Cat\'olica, Av. Vicuña Mackenna 4860, Santiago, Chile.\\
$^{4}$Centro de Astro-Ingenier\'ia, Pontificia Universidad Cat\'olica de Chile, Av. Vicu\~na Mackenna 4860, Santiago, Chile.\\
$^{5}$Instituto de Astronom\'{\i}a y F\'{\i}sica del Espacio,
CONICET-UBA, Casilla de Correos 67, Suc. 28, C1428ZAA, Ciudad
Aut\'onoma de Buenos Aires, Argentina.\\
$^{6}$International Centre for Radio Astronomy Research (ICRAR), M468, University of Western Australia, 35 Stirling Hwy, Crawley, WA 6009, Australia.\\
$^{7}$Australian Research Council Centre of Excellence for All-sky Astrophysics (CAASTRO), 44 Rosehill Street Redfern, NSW2016, Australia.
}

\date{Accepted 2022 May 27. Received 2022 May 26; in original form 2021 June 10}

\pubyear{2021}

\begin{document}
\label{firstpage}
\pagerange{\pageref{firstpage}--\pageref{lastpage}}
\maketitle

\begin{abstract}
Observations show that the  surface brightness  of disc galaxies can be well-described by a single exponential (TI), up-bending (TIII) or down-bending (TII) profiles in the outskirts. Here we characterize the mass surface densities of simulated late-type galaxies from the  {\sc eagle} project according to their distribution of mono-age stellar populations, the star formation activity and angular momentum content.
We find a clear correlation between the inner scale-lengths and the stellar spin parameter, $\lambda$, for all three disc types with $\lambda>0.35$. The outer scale-lengths  of TII and TIII discs  show a positive trend with $\lambda$, albeit weaker for the latter. TII discs prefer fast rotating galaxies. 
With regards to the stellar age distribution, negative and U-shape age profiles are the most common  for all disc types. Positive age profiles are determined by a more significant contributions of young stars in the central regions, which decrease rapidly in the outer parts. TII discs prefer relative higher  contributions of old stars  compared to  other mono-age populations across the discs whereas TIII discs become progressively more dominated by  intermediate age (2-6 Gyrs) stars for increasing radius. The change in slope of the age profiles is located after the break of the mass surface density. We find evidence of larger flaring for the old stellar populations in TIII systems compared to TI and TII, which could indicate the action of other processes. Overall, the relative distributions of mono-age stellar populations and  the dependence of the star formation activity on radius are found to shape the different disc types and age profiles.

%{Observations show that the  surface brightness of disc galaxies  can be well-described by a single exponential profile (TI) while others show a break with either up-bending (TIII) or down-bending (TII) in the outskirts. The origin of these different profiles is still an open question. 
%We use the  {\sc eagle} cosmological hydrodynamic simulation to build a sample of well-resolved disc galaxies. We explore the characteristics of the mass surface densities according to the distribution of mono-age stellar populations, the star formation activity and the angular momentum content.

%We find that the inner scale-lengths of TII correlate with the galaxy stellar spin parameter $\lambda$, while those of TI and TIII show a correlation only for $\lambda>0.35$ and an anticorrelation for lower $\lambda$ values. The outer scale-lengths  of TII discs also show a positive trend with $\lambda$ whereas TIII discs show a weak dependence.
%We find evidence of larger flaring for old stellar populations in TI and TIII systems. Discs have different stellar age profiles: negative, U-shape, positive and $\Lambda$-shape profiles. The negative and U-shape age profiles are the most common ones {\bf for all disc type.} %while positive and $\Lambda$-shape distributions are detected preferentially in TI discs.%The frequency of these later two profiles can provide constraints on the subgrid physics. 
%The relative distributions of mono-age stellar populations {\bf and } the dependence of star formation on radius {\bf modulate } the different disc types and age profiles. }

\end{abstract}


%%%%%%%%%%%%%%%%%%%%%%%%%%%%%%%%%%%%%%%%%%%%%%%%%%

%%%%%%%%%%%%%%%%% BODY OF PAPER %%%%%%%%%%%%%%%%%%

\section{Introduction}
\label{sec:intro}

Stellar populations store important information on the evolution of galaxies and the physical processes that shaped their morphologies, star formation histories and chemical properties, among others. In the case of spiral galaxies, powerful  information can be readily obtained through their surface brightness (SB) profiles. Disc components  are typically consistent with well-behaved exponential  profiles \citep{patterson1940,freeman1970}. To first order, their formation can be explained by  the so-called standard model for disc formation \citep{fallyefs1980} which is based on the specific angular momentum conservation of baryons. However, within the current cosmological paradigm, galaxy formation proceeds in a non-linear way so that mergers and interactions with other galaxies, or with the global environment,  can disturb galactic discs causing
a redistribution of angular momentum \citep[e.g.][]{pedrosa2015,teklu2015,Lagos2017}, the formation of non-asymmetric features \citep[e.g.][]{FGomez2016,FGomez2017,FGomez2021,grand2016} and the loss of material via tidal stripping \citep{bh96}, among others. 
It has become clear that not all spiral galaxies can be described by a single exponential profile extended to arbitrarily large radii  \citep{kruit1979}.  Thus, while the SB distributions of some discs  can be well-described by an exponential profile all the way out (TI), other discs show  a brightness deficiency (TII) or excess (TIII) with respect to a single exponential profile. Such discs are are better described by  double exponential fits \citep[e.g.][]{potru2006} with a characteristic break radius (\rbreak). Observational studies suggest a correlation between the Hubble type and the disc type. TIII systems are reported to be more  frequent  in early-type spirals while  TII ones  are more common in late-type ones \citep{erwin2008,potru2006,gutierrez2011}. 

Several theoretical and observational works have analysed  the  characteristics of outer regions of disc galaxies \citep[e.g][]{roskar2008a,roskar2008b,sabla2009,minchev2012,bakos2008,yoachim2012,rula2016}. Yet, the origin of the deviation from a single exponential disc is still unclear. Variations of star formation efficiency, migration, accretion of small satellites, or a combination of all of them could be behind this. The  main hypotheses formulated to explain "breaks" in the SB profiles are related to  angular momentum redistribution \citep{debattista2006}  and to the existence of a star formation threshold \citep{elhu2006}. Additionally, several works have reported a link between an outer upturn ("U-shape") in the age profiles and  the TII profile, with the age up-turn causing the lack of light in the outer regions \citep{bakos2008,roskar2008b,sabla2009,marse2009,yoachim2012}.
According to \citet{sabla2009}, the origin of TII and TIII galaxies arises from a combination of two processes:  (1) an abrupt change of slope in the radial star formation profile and (2) the effects of stellar radial migration inside-outside of \rbreak. The first is due to a change in the gas volume density profile, which  would cause the bending down in the stellar density profile. The second is due to radial migration of stars formed in the inner parts towards locations beyond \rbreak. In addition, these authors proposed that the different types of stellar density profiles could be a consequence of a variation of efficiencies of both mechanisms. \citet{rula2017}  reported different stellar age and metallicity inner gradients for galaxies displaying  TI, TII and TIII SB profiles. Those results are interpreted as outcome of a gradual increase in the radial migration efficiency of stars from TII to TI and TIII galaxies.

\citet{rula2016}  analysed 44 spiral galaxies selected from the CALIFA survey \citep[][see also \citealt{rula2017}]{califa}. They compared the SB and age profiles in order to find differences between profile types (I and II). They also reported a U-shape in the age profiles of 17 galaxies that were either TII or TI.  However in their sample, the U-shape feature  is only observable in light-weighted age profiles. They claim that the mechanisms shaping the SB and stellar population distributions are not directly coupled and that the U-shape in age profiles would be due to an early formation of the  disc followed by an inside-out quenching of the star formation. 

On the other hand, \citet{herpich2015,herpich2017} investigated the role of the halo spin parameter ($\lambda$) in shaping the outer SB profiles by analysing a set of  simulations of isolated controlled galaxies. They found a clear transition from TIII galaxies, displaying low  spin parameters to TII galaxies, showing higher values. TI discs are reported to have intermediate values, $\lambda \sim 0.035$. Recently, \citet{wang2018} found that the formation of TIII galaxies were linked to  high HI-richness and low spin $\lambda$ of the inner stellar disk. Hence there are still open questions regarding the origin of these characteristics features in the SB profiles. While the smooth formation of discs with global angular momentum conservation would lead to a correlation between scale-lengths and spin parameter $\lambda$ \citep{mmw1998},  these distributions could be  disturbed by variety of physical processes such as stellar migration, dynamical heating, satellite accretion, bar formation, among others.

Clearly, the origin of the spread observed in the properties of galactic disk, such as surface density and brightness profiles, is complex and involves the action of several physical mechanisms. As such, these profiles can provide very rich information regarding the formation and evolution of galactic disc. For this purpose, we analyse the stellar surface density of disc galaxies  selected from the simulation with the largest volume of the EAGLE Project \citep{schaye2015}. The EAGLE simulations have been  shown to  reproduce several  galaxy relations such as  the gas content of galaxies at a
given mass \citep{lagos2015,Bahe2016, crain2017}, the evolution of the galaxy
stellar mass function \citep{furlong2015},  the scale-resolved
metallicity-star formation relation \citep{trayford2019}, the metallicity gradients as a function of stellar mass \citep{tissera2019,Tissera2022} and the azimuthal variation of the metallicity gradients at $z=0$ \citep{solar2020}. Our goal is to characterize the origin of the different surface density and age profiles in a cosmological context. { In this work, we  focus on the analysis of  stellar populations and their age distributions in central disc galaxies.  The stellar surface density  profiles, $\Sigma(r)$, are statistically analysed with the goal of identifying correlation between the distribution of mono-age stellar populations and the star formation activity.}

{ The paper is organized as follows. In Section \ref{Simulated} we describe the main characteristics of the EAGLE simulations, the selected galaxy sample and the algorithm that is be applied to classify the different $\Sigma(r)$. Section \ref{analysis} describes the analysis and results. In Section \ref{conclusions} we discuss and summarize our main findings.}

%%%%%%%%%%%%%%%%%%%%%%%%%%%%%%%%%%%%%%%%%%%%%%%%%%%%%%%%%%%

\section{Simulated galaxies}
\label{Simulated}
%\subsection{The EAGLE project}
For this work we will use  largest  simulation (Ref-L100N1504) of the EAGLE Project\footnote{ In http://eaglesim.org,
http://eagle.strw.leidenuniv.nl  a global description of the project as well as access to movies and images and the database of galaxies can be found \citep{mcalpine2016}.} \citep{schaye2015,crain2015}.
The initial conditions are consistent with  the Planck Cosmology parameters  \citep{planck2014}: $\Omega_\Lambda=0.693$, $\Omega_{\rm m}=0.307$, $\Omega_{\rm b}=0.04825$, $\sigma_8=0.8288$, $h=0.6777$, n$_{s}=0.9611$ and $Y=0.248$ where  $\Omega_\Lambda$, $\Omega_{\rm m}$ and  $\Omega_{\rm b}$ are the average densities of dark energy, matter and baryonic matter in units  of the critical density at $z=0$,  $\sigma_8$ is the square root of the linear variance of the matter distribution when smoothed with a top-hat filter of radius $8h^{-1}$ cMpc,  $h$ is the Hubble parameter ($H_{o}\equiv h \,100 \rm km  s^{-1}$), $n_{s}$  is the scalar power-law index of the power spectrum of primordial  perturbations, and $Y$ is  the primordial mass fraction of  helium.
 The subgrid physics parameters were calibrated to reproduce the galaxy mass function and the observed sizes of galaxies  at $z=0.1$  \citep{schaye2015}.  In addition,  other variations of the subgrid physics were explored  as presented in  \cite{crain2015}.

The Ref-L100N1504 run simulates  a cubic volume of side 100 Mpc.  The setup of the initial conditions provides a mass resolution of  $9.7\times 10^6 M_{\odot}$ for dark matter and  an initial mass of $1.81 \times 10^6 M_{\odot}$ for baryonic particles. The gravitational calculations between particles are computed with  a Plummer equivalent softening length of 2.66  comoving kpc limited to a maximum physical size of 0.70 pkpc ($\epsilon_{\rm grav}$).
This simulation provides a large number of galaxies formed in a variety of environments.
%\subsection{The galaxy sample}

\begin{figure}
    
  \includegraphics[width=0.9\columnwidth]{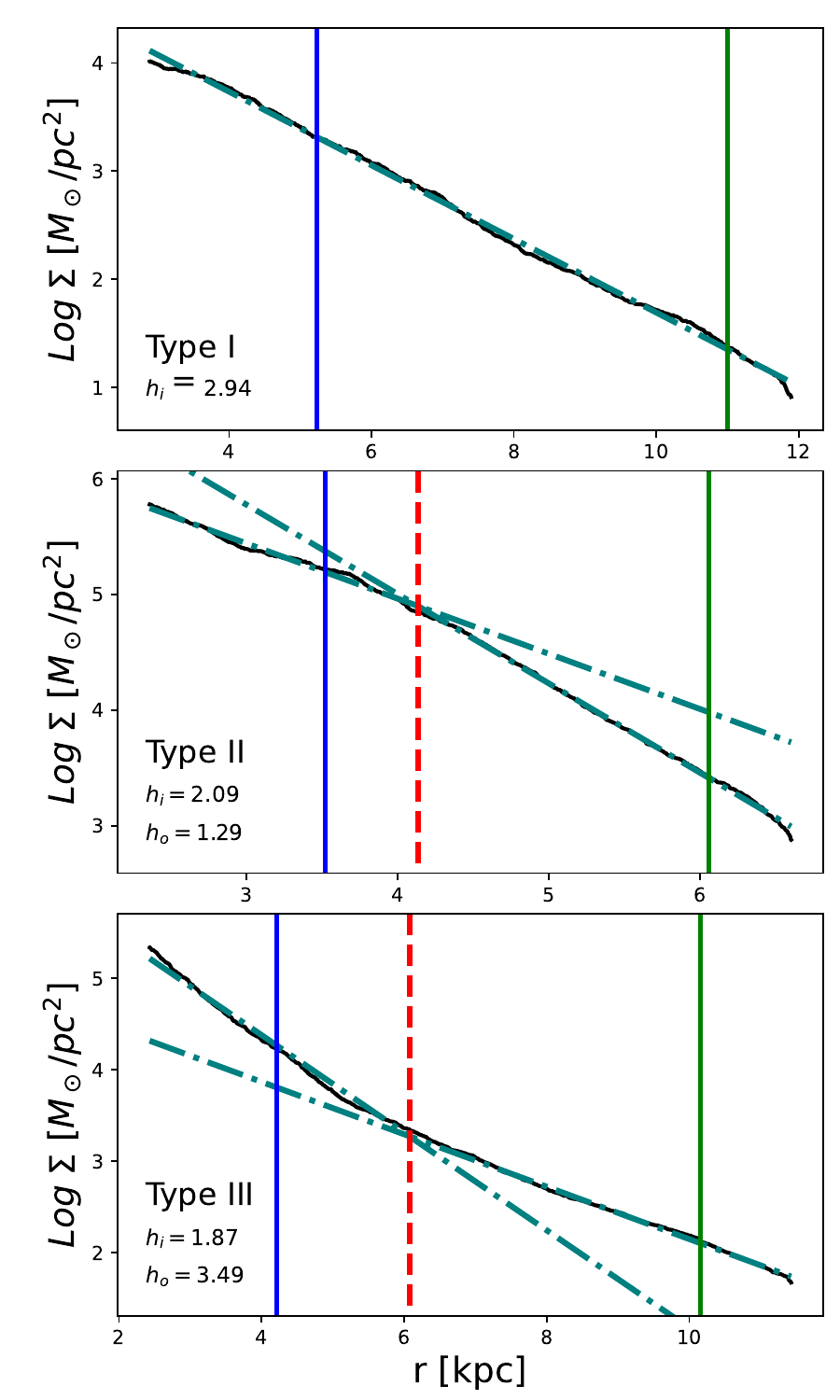}
\caption{Examples of  typical 
  TI (upper panel), TII (middle panel) and TIII (lower panel) $\Sigma(r)$ profiles in the EAGLE discs (black lines). The linear regression for TI and the double exponential fits for TII and TIII are also shown (cyan, dashed dot lines).  The characteristics scales, $R_{\rm eff}$ (blue, solid lines), $ R_{\rm opt}$ (green, solid lines) and \rbreak (red, dashed vertical lines) are also depicted, together with the 
  inner $h_{\rm i}$ and outer $h_{\rm o}$ scale-lengths.}
\label{examtypes}
\end{figure}

\subsection{The simulated galaxy sample}
\label{subsimulated_galaxy_sample}
We use the galaxy catalogue constructed by \citet{tissera2019} from the Ref-L100N1504 at z=0. Only central galaxies within a given halo are analysed (i.e. no satellite galaxies have been included).
%The simulated galaxies were identified by using a Friends-of-Friends algorithm to select virialized structures and then the {\small SUBFIND} scheme \citep{springel2001} to extract the main galaxies. 
The spheroidal and disc components were separated by applying a dynamical criterion based on the angular momentum content and the binding energy. This is done by estimating $\epsilon = J_{\rm z} /J_{\rm z,max}(E)$, where $J_{\rm z}$ is the angular momentum component in the direction of the total angular momentum, and $J_{\rm z,max}(E)$ is the maximum $J_{\rm z}$ over all particles at a given binding energy $E$ 
\citep[see][for details on the procedure and conditions used]{tissera2012}. With this algorithm, the disc and bulge components are identified.
The disc-to-total stellar mass ratio ($D/T$) is used to classify galaxies
according to their morphologies.
For this analysis, central galaxies with $D/T > 0.5$ are selected. Additionally, only those  with disc components resolved with more than 5000
stellar particles are considered. As a consequence, our sample comprises 1012 disc galaxies with stellar mass within the mass range [$10^{10}$, $10^{11.5}$] $\rm M_\odot$. For each galaxy we calculated the optical radius, \ropt,  as  the radius that enclosed $\sim$83 percent of the stellar mass of a disk \citep{saiz2001}, and the effective radius, \reff,  as the radius that enclosed half of the disc stellar mass.
 
 Hereafter, the $\Sigma(r)$ profiles of the disc components of the selected galaxies are defined by projecting the stellar mass onto the rotational plane of the discs{, from $3\epsilon_{\rm grav}$ to $1.5\; R_{\rm opt}$}. This is performed after the galaxies are rotated to have their z-axis aligned with the total angular momentum of the stellar discs.
%To characterize the outskirt of the simulated galaxies, we build

To minimize potential issues due to numerical resolution or very local inhomogeneities, the $\Sigma(r)$ profiles are calculated  using moving averages. In practice, for a given galaxy, the azimuthally averaged disc surface density, $\Sigma(r_i)$, is estimated at the galactocentric distance of each disc stellar particle, $r_i$. Here $i=1, ..., N_{\rm disc}$ where $N_{\rm disc}$ is the total number of disc particles. To do so, we first sort all disc stellar particles with respect to $r_i$ in ascending order. We  compute  $\Sigma(r_i)$ by selecting both the 10 percent of disc particles that are located at $r > r_i$ and the 10 percent of disc particles that are located at $r < r_i$. In other words, a total of 20 percent of disc stellar particles are considered at each $r_i$. We then add the mass of all selected disc stellar particles and compute the surface density considering the area enclosed by the selected subset\footnote{Note that, to always use the same number of particles when estimating $\Sigma(r_i)$, for the first and last 10 percent of the sorted disc particles, $\Sigma(r_i)$ is not computed.}.

\begin{figure*}
    \begin{center}
        %\plotone{Images/Repre/Age_DISK_STAR.pdf}
        \includegraphics[width=0.75\textwidth]{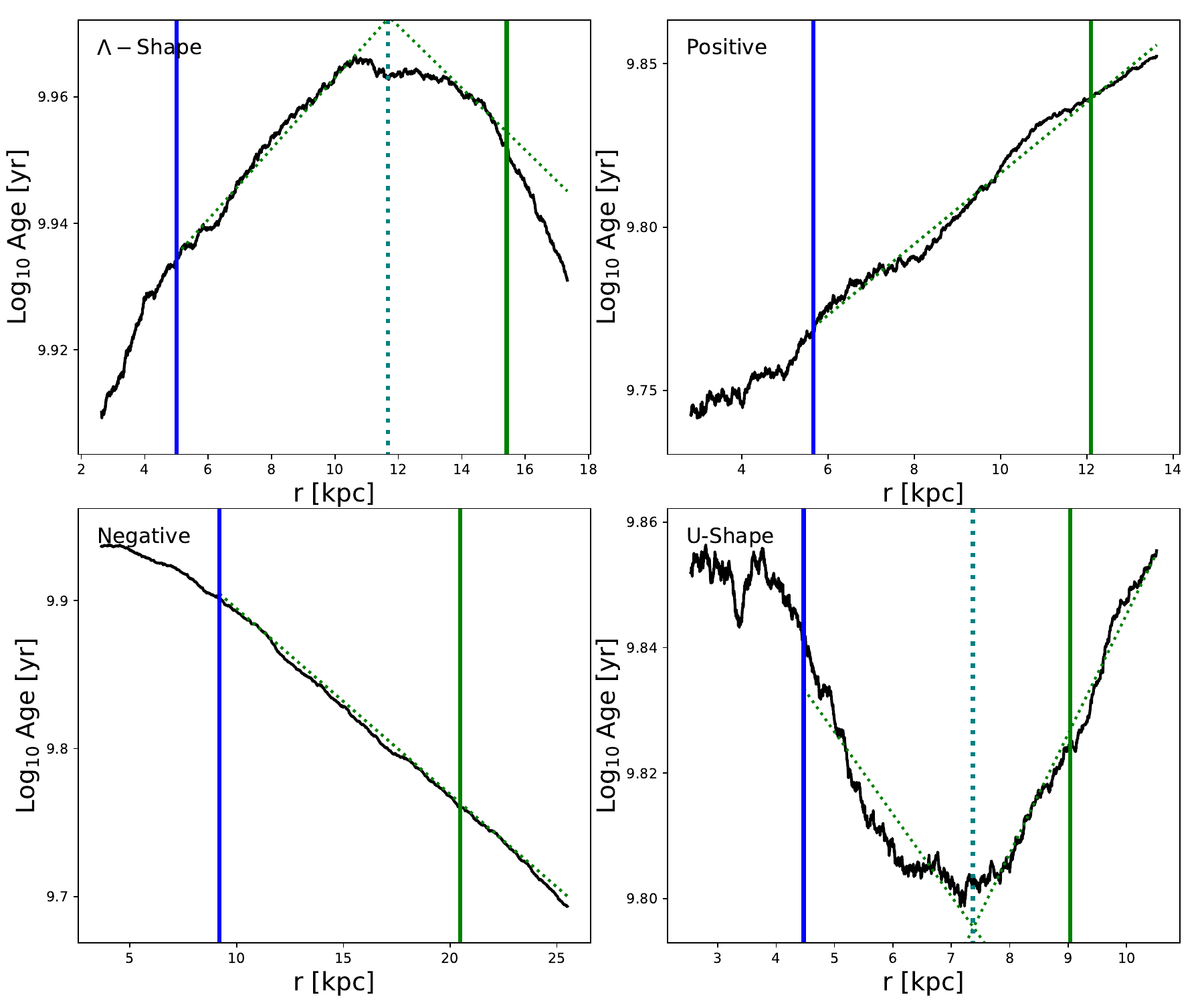}
    \end{center}
  
    \caption{Four different age profiles identified in EAGLE discs are shown for illustration purposes (black lines). The 
linear regressions are also depicted (green dotted lines). The characteristics scales, $R_{\rm eff}$ (blue solid lines), $ R_{\rm opt}$ (green solid lines) are included. Additionally,  for the U-shape and $\Lambda$-shape profiles $R_{\rm min}$ and $R_{\rm max}$ are depicted (green short-dashed lines). }
\label{pefilesages}
\end{figure*}

\subsection{Break-Finder (BF) algorithm}
\label{BFAlg}

In order to characterize the $\Sigma(r)$ profiles, we follow a methodology similar to that established by previous authors \citep{erwin2008,mateo}, where $\Sigma(r)$ profiles are presented as a double power-law piece-wise joined by an inflection point \citep[see][equations 5 and 6]{erwin2008}. For this, we fit the profiles using the following piece-wise function:
%we perform  a double power-law fit joined . :
\begin{equation}\label{eqfit}
 \log_e \Sigma_{\rm i} (r) = \log_e \Sigma_{0,\rm i} - \frac{r}{h_{\rm i}} \quad  if \quad r \leqslant R_{\rm br},
\end{equation}
\begin{equation}\label{eqfit2}
 \log_e \Sigma_{\rm o} (r) = \log_e \Sigma_{\rm 0,o} - \frac{r}{h_{\rm o}} \quad if \quad r>R_{\rm br},
\end{equation}

where the two parts are joined at the break radius, \rbreak. Here $h_{\rm i}$ and $h_{\rm o}$; inner and outer scale lengths, $\Sigma_{0,\rm i}$ and $\Sigma_{0,\rm o}$; inner and outer central densities and \rbreak. Note that, the value of  $\Sigma_{0,\rm o}$ is set  by the values of the remaining four parameters. Equations \ref{eqfit} and \ref{eqfit2} are used to characterize the inner and outer regions of the discs, respectively.

Our BF fitting procedure considers star particles within 0.5 $R_{\rm eff}$  to $1.5 R_{\rm opt}$. The BF algorithm is sensitive to breaks. A potential problem with the methods arises at the disc outer and inner most edges, where spurious inflection point could be defined. To avoid spurious results we allow \rbreak to only vary within $0.5 ~ R_{\rm eff}$ to $R_{\rm opt}$. The remaining parameters are forced to be greater than 0. The parameter fitting is performed using a robust least squares optimization method for bounded problems called Trust Region Reflective algorithm \footnote{ We implement Trust Region Reflective algorithm using the {\sc curve\_fit} function from the {\sc scipy} library. More details can be found at https://scipy.org}. The best fit minimizes the sum of squared residuals, which is quantified through the $\chi^2$. The result is always a double exponential. The type of profiles are defined according to the criteria described in section \ref{SigmaSection}

\section{Analysis}
\label{analysis}
\begin{figure*}
    \begin{center}
 \includegraphics[width=0.75\textwidth]{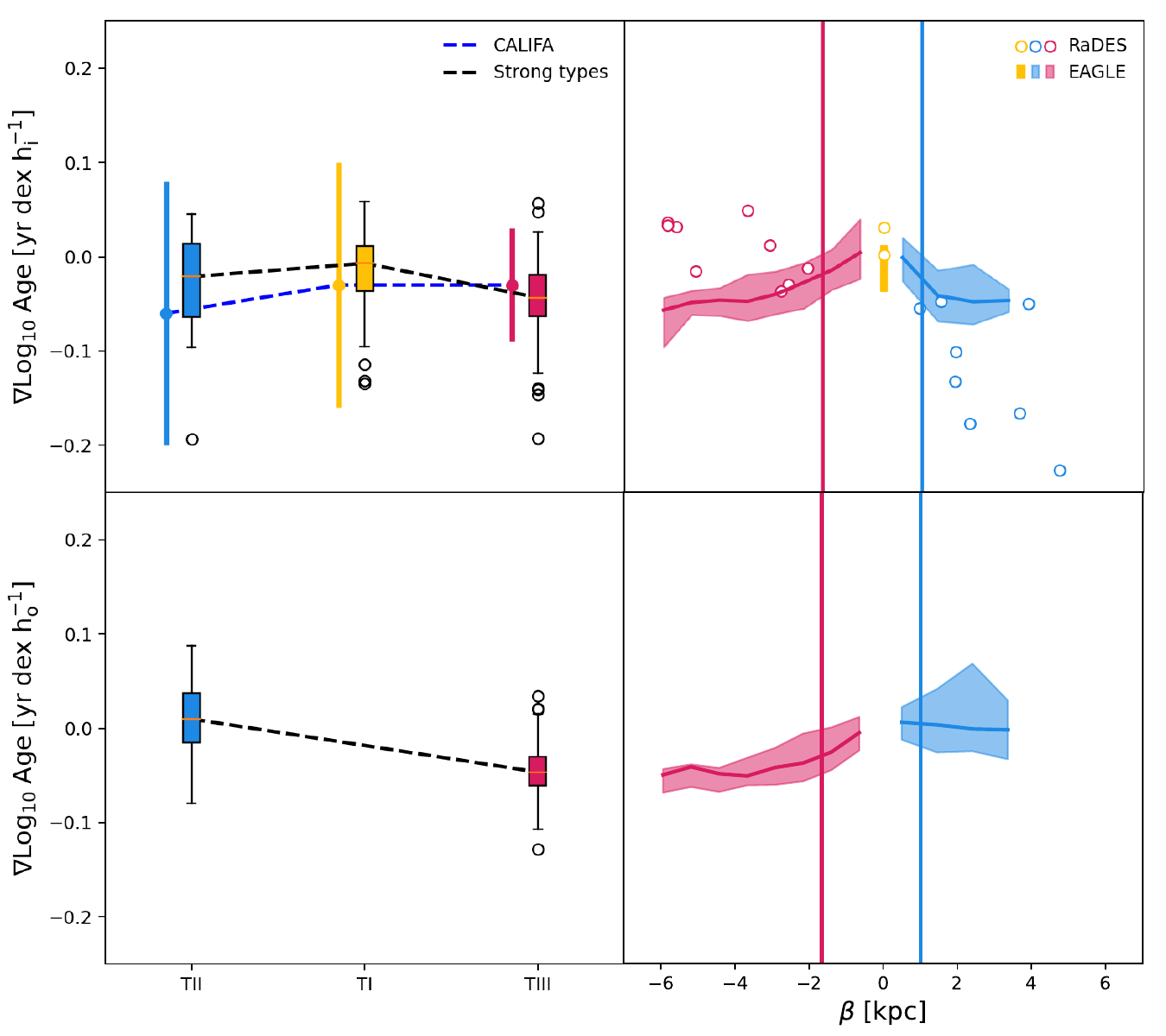}
 \end{center}
  \caption{Analysis of the gradients of the inner and outer age
   profiles for different disc types (top and bottom panels, respectively).   
   In the left panels, the age gradients for the inner (upper) and outer (lower) discs  as a function of disc types for our EAGLE sample are shown. The median age gradients for the strong  types subsamples (blue, yellow and red boxes for TII, TI and TIII, respectively and dashed black lines) are included. The size of the boxes denote the $25^{\rm th}$ and $75^{\rm th}$ percentiles (open black circles show the outliers). For comparison the median mass-weighted ages gradients  for the inner discs  reported  by  \citet{rula2017} for galaxies in the CALIFA survey (dashed blue line) are also depicted (the error bars correspond to the dispersion as given by their table 4).
   In the right panels, the median age gradients for the inner and outer discs are shown for TI, TII and TIII galaxies. The shaded regions are enclosed by the  $25-75^{\rm th}$ percentiles of the corresponding distributions. The red and blue vertical lines depict the medians of $\beta$ for the TIII and TII profiles, respectively. These are used to define strong TII and TIII types (Sec. \ref{SigmaSection}). The results of \citet{rula2017a} using the RaDES simulated galaxies (red and blue open points) are also included).}
\label{ruizlara}
\end{figure*}

\subsection{Stellar surface density  profiles}
\label{SigmaSection}

We analyse  the $\Sigma(r)$  of the selected disc galaxies and classify them as TI, TII and TIII profiles. For this purpose, the BF algorithm described in Section \ref{BFAlg} is applied to the $\Sigma(r)$  of all galaxies in our sample.  The classification is performed by comparing the $h_i$ and $h_o$ scales. The strength of the break is quantified by defining
\begin{equation*}
\centering
\beta = h_i - h_o.   
\end{equation*}
Following \citet{rula2017}, we define TII discs are those with  $\beta \gtrsim 0.5$ {kpc}, TIII discs  have  $\beta \lesssim -0.5$ {kpc}, and TI discs have with $|\beta| \leq 0.5$ {kpc}. This  value corresponds to the quarter of $\beta$ standard deviation. Then, for TI discs, we perform a single power-law fit. In Fig.~\ref{examtypes}, we show the result of applying the fitting algorithm to three typical examples.
Clearly, TII and TIII discs show  double exponential profiles, with the inner disc defined for $r< R_{\rm br}$ (the cyan, dashed vertical line), and the outer disc for $ r > R_{\rm br}$.

Finally, we use the $\chi^2$ of the best fits to eliminate from the samples those discs that have more complex structure and hence require more than two exponentials to describe the  $\Sigma(r)$. These discs are
strongly disturbed and are not suitable for the study of this paper.
For this purpose, 
 only galaxies with  $\chi^2 < \hat{\chi}^2 + \sigma_{\chi^2}$, where $\hat{\chi}^2$ is the median and $\sigma_{\chi^2}$ is the standard deviation of the $\chi^2$ distribution for our total sample. 
  In this way, we eliminate the outliers of $\chi^2$ at $1-\sigma$ level, and build a well-fitted subsample of simulated galaxies whose $\Sigma(r)$ can be described by either a pure  or a double exponential. Our final sample comprises 912  galaxies.

%For this purpose, 
% only galaxies with  $\chi^2 \lesssim 4.56$ are considered. This  reference value is estimated as $<\chi^2> + \sigma_{\chi^2}$, where $\sigma_{\chi^2}$ is the  standard deviation of the $\chi^2$ distribution. 
%  In this way, we eliminate the outliers of $\chi^2$ at $1-\sigma$ level, and build a well-fitted subsample of simulated galaxies. {\bf Finally, our sample are reduced to 912 well-fitted galaxies.}

With the aim of analysing a more trustworthy sample, we quantify the strong TII and TIII types groups by selecting the fifty percent of galaxies of each $\Sigma$ type which have the strongest deviation between the inner and outer discs. For this purpose, the  median of the $\beta$ distribution for each $\Sigma$ type is
calculated.  Strong TII are  galaxies that have  $\beta$ larger than $\hat{\beta} = $1.02 {kpc}, while  strong TIII galaxies have $\beta$ less than $\hat{\beta}=$-1.66 {kpc}. Of course, there are no strong types for TI discs. 
In Table~\ref{tablesamples}, we show the numbers of members and percentages of galaxy types in the total and strong samples. We can see that TIII  is the most frequent type overall in the analysed sample. However TI are more frequent if we only consider the strong TII and TIII subsamples. The less frequent galaxy type in both EAGLE samples is TII. Hereafter, we will only consider the  strong-type discs in the analysis with the purpose of disentangling clearly the similarities and differences among them.

%Where this last values corresponds to the median of $\beta$ of between TII and TIII of our sample, respectively.

As can be seen from Table~\ref{tablesamples}, TI and TIII are the most 
frequent types  in our simulated sample. This seems at odd with  observations, which find  TII to be the most common profile. 
However, observations also suggest that the fractions of galaxies with different disc types vary with morphology and probably with the environment. TII discs are reported to be more frequent in late-type galaxies while TIII discs tend to be identified in early-types. For example, \citet{potru2006}  analysed a sample of 93 late-type galaxies and found a distribution of 10\%:60\%:30\% for types TI:TII:TIII, respectively.  Using 66 early-type spiral galaxies, \cite{erwin2008} reported  percentiles of  27\%:42\%:27\%, and \cite{gutierrez2011} also detected a trend for TII to be more frequent in late-type spirals using a sample of 47 face-on early-type unbarred galaxies, they reported a global breakdown of 21\%:50\%:38\%. As it will be discussed in  Sec. \ref{SecAngularMom}, this dependence on morphology is reproduced by the EAGLE simulations. Although the frequencies between our EAGLE sample and the observations are not similar, recent work with galaxy triplets finds a high frequency of TIII galaxies in this environment. Members of such triplets typically show signs of interactions, highlighting the relevance of environment \citep{Tawfeek2021}. We note that, in our work, galaxies are located in different environments \citep{tissera2019} and that this aspect will be addressed in  a forthcoming paper.

%{\bf In order to compare this results with observations, \cite{potru2006} from a sample of 93 late-type galaxies, found a distribution of 10\%:60\%:30\% for types TI:TII:TIII. On the other hand, \cite{erwin2008} using 66 early-type galaxies, found a proportion of 27\%:42\%:27\%, and \cite{gutierrez2011} found 21\%:50\%:38\%, with a sample of 47 face-on early-type unbarred galaxies. Interestingly, in these last two works, they found that TII profiles tend to be late-type galaxies, while TI and TIII tend to be early-types (we will mention about this in Sec.\ref{SecAngularMom} ).  }

\begin{table}
   \begin{center}
   \caption{Distribution of the disc types in the {\sc EAGLE} sample.}
 \label{tablesamples}
 \begin{tabular}{lrcrrr}\hline
   &N$_{\rm T}$&$ N_{\rm TI}$ & $N_{\rm TII}$ & $N_{\rm TIII}$ \\
 \hline
All Types &912& 301 (33\%) & 116 (13\%)& 495 (54\%)\\
Strong Types & 607 & 301 (49\%) & 58 (10\%) & 248 (41\%)\\
 \end{tabular}
  \end{center}
 \end{table}
 
\begin{figure}
\begin{center}

\includegraphics[trim={0.5cm 0cm 1.5cm 1.5cm},clip,width=0.9\columnwidth]{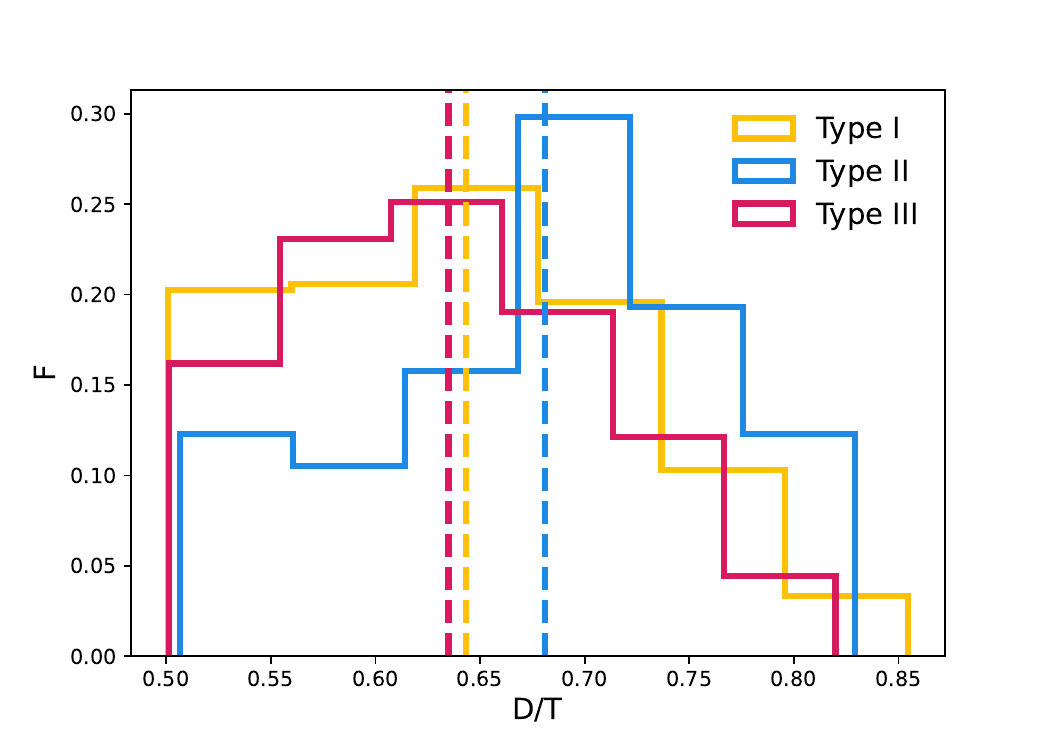}
  \includegraphics[trim={0.5cm 0cm 1.5cm 1.5cm},clip,width=0.9\columnwidth]{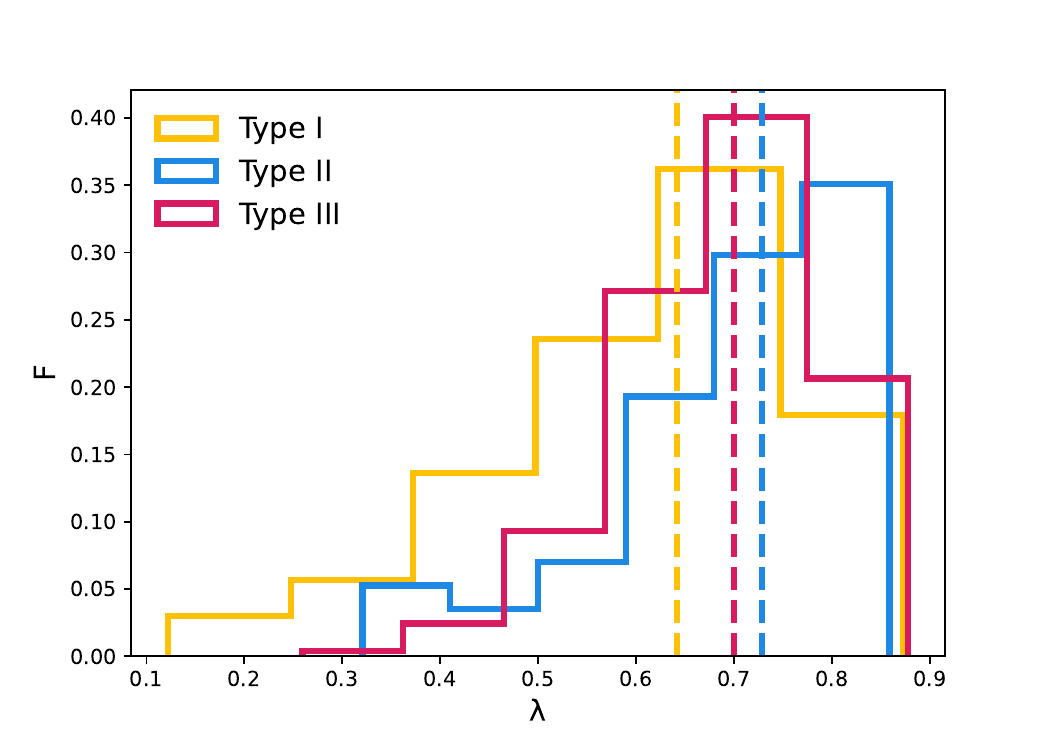}
  \includegraphics[trim={0.5cm 0.5cm 1.5cm 1.5cm},clip,width=0.9\columnwidth]{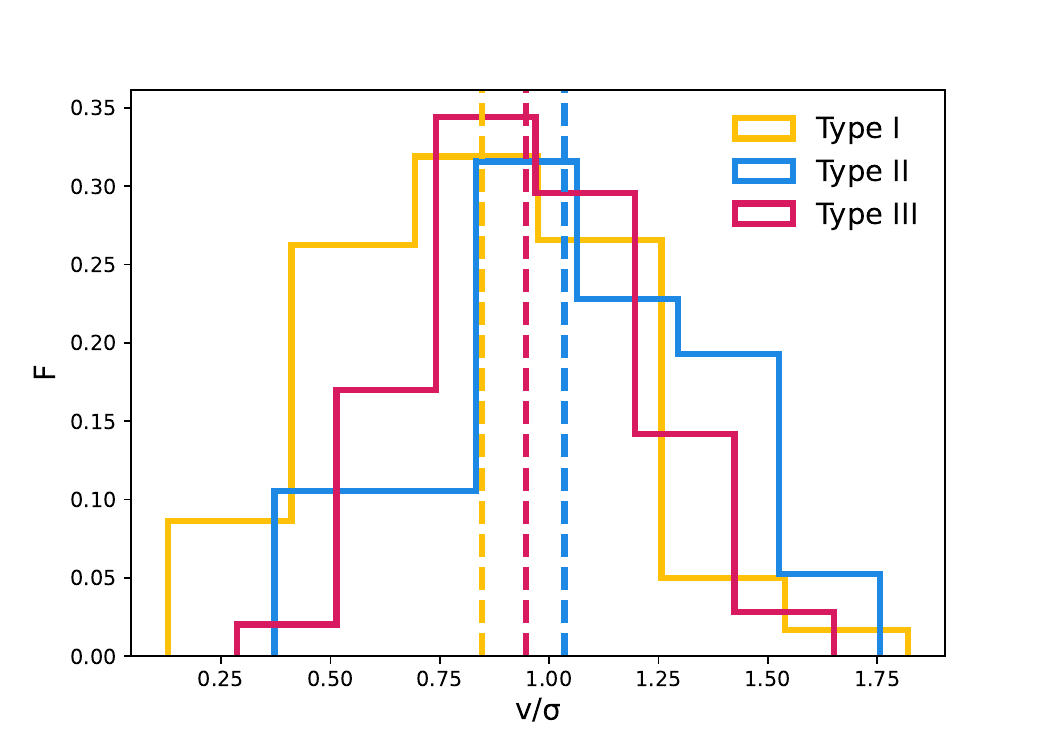}
  %\plottree
\caption{Distribution of galaxies with TI, TII and TIII discs as a function  of  $D/T$ (upper panel), $\lambda$ (median panel) and $v/\sigma$ (lower panel).  The median values for each sub-sample are  shown as indicated in the legend. The histograms are normalized by the total number of elements in each subsample.}
\label{figlambda}
\end{center}
\end{figure}

\subsection{Age profiles}

To understand the origin of the different types of $\Sigma(r)$ profiles, and to confront them with
observations,  we also estimate  age profiles for each galaxy.
%and oxygen abundance profiles.
%Figure \ref{pefilesages} shows examples of the
%diversity of age profiles found in the EAGLE discs.
As for the $\Sigma(r)$ profiles, the mass-weighted stellar age profiles are constructed by using stars located between $3\epsilon_{\rm grav}$ and $1.5\; R_{\rm opt}$ and are estimated by using moving averages. {We note that the age profiles have been calculated within the same radial range used for the
surface density profiles in order to be able to correlate their trends.}
%\begin{equation}
%\langle age (yr)\rangle _{M-W} =  \frac{\sum_{i} age_i * mass_i} {\sum_i mass_i} 
%\label{eqage}
%\end{equation}
%where i is the position of a star using moving averages. 
%Then a double-linear regression is fitted to the profile following the procedure mentioned in Section \ref{BFAlg}
 Then  a double-linear piece-wise function is fitted within the radial range [\reff,1.5\ropt]\footnote{The radial range have been extend towards larger radii to ensure the detection of  any change of slope in the age profiles since they are found to occur at larger galactocentric distances than the \rbreak.} using our BF algorithm. {The profiles were fitted using logarithmic scale}. In this way the inner, $a_{\rm in}$,  and outer, $a_{\rm out}$, slopes  %($a_{in}$  and $a_{out}$, respectively)  
 and the inflection point are obtained (see Section \ref{BFAlg}). To classify the age profiles, we use the $a_{\rm in}$ and $a_{\rm out}$ slopes. We define four different behaviours by applying the criteria described below:
{
\[ a_{\rm in}<0 \quad
  \begin{cases}
    \text{U-shape}   & \quad  \text{if }\; a_{\rm in}<a_{\rm out}\\
    \text{Negative}   & \quad  \text{if }\; a_{\rm in} \geqslant a_{\rm out}\\
  \end{cases}
\]
\[ a_{\rm in}>0 \quad 
  \begin{cases}
    \text{Positive}   & \quad  \text{if }\; a_{\rm in} \leqslant a_{\rm out}\\
    \text{$\Lambda$-shape}  & \quad  \text{if }\; a_{in}>a_{\rm out}\\
  \end{cases}
\]
}

For the U-shape profiles, we refer the radial position at which the profiles with slopes $a_{\rm in}$ and $a_{\rm out}$ intersect each other as $R_{\rm min}$. For the $\Lambda$-shape profiles this radial position is referred as $R_{\rm max}$. {These definitions are related to the position of the minimum and maximum age, respectively}. Fig.~\ref{pefilesages} show typical examples of positive, negative U- and $\Lambda$-shape age profiles. 
These different types of age profiles are found in simulated discs regardless of their $\Sigma(r)$ type (i.e. TI, TII and TIII). However, their frequencies change for TI, TII and TIII discs as summarized in Table 2. On the one hand, the U-shape profile is found to be the dominant class  in TI and TII discs %, and, the second most frequent is the negative age profiles. 
while for TIII discs the most frequent are negative and U-shaped age profiles. On the other hand, positive and $\rm \Lambda$-shape age profiles are the least frequent in TIII galaxies (where together represent only 4 percent to the total TIII), but are more numerous in TI where they are found in $\sim$37 per cent of the total TI galaxies.
Observationally, \citet{Roediger2012} reported 15\% of the TI and 36\% of TII and TIII to have U-shape profiles. They also reported 30\% of TI to have negative age slopes. Our estimations for the {\sc EAGLE} galaxies are in general agreement with these observations. \citet{Roediger2012} also found that more than 50\% of galaxies in their Virgo sample had positive age gradients regardless of the disc types, indicating the possible action of environmental effects. However, we have not found any report  of $\Lambda$-shape age profiles in observations yet. 
%ORIGINAL
%However, their frequencies change for  TI, TII and TIII disks,  as summarized in Table~\ref{tabage}.  U-shape  profile is the dominant class for the age profiles in TI and TII discs. For those, the second most frequent is the negative age profiles, however is dominant for TIII . Positive and $\Lambda$-shape age profiles are the less frequent ones, particularly in TIII  discs. We found that TI and TII discs have a similar frequency of $\Lambda$-shape age profiles,  accounting for about 20 and 10 per cent of the samples, respectively. Interestingly, only $\sim 5$ per cent of TIII discs show positive or $\Lambda$-shapes  age profiles. 

%%%%%%%%%%%%%%%%%%%%%%%%%%%%%%%%%%%%%%%%%%%%%%%%
 \begin{table}
   \begin{center}
   \caption{{Age profile type frequencies in TI, TII and TIII discs. The fractions of galaxies with a given age profiles in a given $\Sigma$ type  subsample are provided within parenthesis. }}
 \label{tabage}
 \begin{tabular}{lrrr}\hline
   & TI & TII & TIII \\

 \hline
Negative        & 85 (0.29)&     13 (0.22) &     120 (0.50)\\
 U-shape        & 101 (0.34) &     25 (0.43) &     110 (0.46)\\
 Positive        & 49 (0.17) &     14 (0.24) &     6 (0.02)\\
 $\Lambda$-shape & 59 (0.20) &     6 (0.10) &     5 (0.02) \\

 \end{tabular}
  \end{center}
 \vspace{1mm}
 \end{table}

%Note that in order to stack of profiles, work with normalized profiles by \rbreak for TII and TIII profiles, but in the TI galaxies are normalized by $R_{\rm opt}$ (for SFR and $\Sigma$ distributions), but in Fractional mass and <|Z|> distributions work with normalized by \ropt.

%\subsubsection{Comparison with observations}
 We compare our results with  \citet{rula2017a}, where  214 spiral galaxies selected from the CALIFA survey are analysed. {They estimated both the light-weighted and mass-weighted age profiles. To make a fair comparison with our simulated data, we adopted the  slopes of the linear regression fits of their mass-weighted age profiles. }We also consider the definitions proposed by \citet{rula2017} to compute the inner and outer gradients.
 The inner discs are defined between  0.5$\mathrm{R_{\rm eff}}$ and $\mathrm{R_{\rm br}}$  for TII and TIII discs, and between 0.5$\mathrm{R_{\rm eff}}$ and 1.5$\mathrm{R_{\rm opt}}$ for TI discs. The outer regions in TII and TIII discs are calculated form  $\mathrm{R_{br}}$ to  1.5$\mathrm{R_{opt}}$. The inner gradients are normalized by the inner scale-length $h_i$ and $h_o$ for outer gradients.
 
 %The linear regression are estimated in the range[0.5$\mathrm{R_{eff}}$, $\mathrm{R_{br}}$] for TII and TIII discs, but in the TI disc, the profiles are fitted between 0.5$\mathrm{R_{eff}}$ to $\mathrm{R_{opt}}$. The gradients are normalized by the inner scale-length $h_i$.
 
 %, obtain various types of profiles such as colour, age, metallicity and surface brightness profiles, to characterize their light distributions. 
 In the left panel of Fig.~\ref{ruizlara}, we display the age gradients for TI, TII and TIII, including the observational results presented in \citet{rula2017}. Our sample is represented by the {colored boxes} in the left panels.
 As can be seen, the simulated age  profiles  for each $\Sigma(r)$ type are within the observed range. 
 However, there is a trend for the simulated age gradients to be slightly shallower for 
TI and TII.
%agreement with the results recently reported by
%\citet{tissera2019}. 
%transformation of star-forming gas into stars at all radii drives the
%weak metallicity profiles.
%Because of the general flatter slopes, there are clear trend with disc types.
In the top right panel of  Fig.~\ref{ruizlara} the median age gradients for the strong TII and TIII for our EAGLE sample are displayed as a function of $\beta$  and compared with the results of RaDES  simulated discs \citep[][open circles]{rula2017}. {We note that these authors estimated light-weighed age profiles using the r-band SDSS luminosities and this might introduce difference with our results. }
The EAGLE discs show a weak correlation for TIII and anticorrelation for TII between the inner age gradients with $\beta$. The Pearson correlation factor of each one are $0.37$ and $-0.30$ respectively. Regarding TIII discs, RaDES reported an anticorrelation while EAGLE is consistent with the opposite trend. These differences could originate by the different subgrid physics implemented in RaDES and EAGLE simulations. Hence, the information stored in both age and $\Sigma$ profiles could help to constrain the galaxy formation models.

In the lower panels of Fig.~\ref{ruizlara} we also include similar relations for the age gradients in the outer regions using $h_o$. This plot can be done only for TII and TIII discs. As can be seen the age profiles are shallower/positive for TII, while the outer regions of TIII discs show stronger negative age profiles and a trend for stronger TII to have more negative outer age gradients.
%The $\beta$ parameters show consistent trends.
%The latter agrees well with the trend reported by \citet{rula2017}.

%%%%%%%%%%%%%%%%%%%%%%%%%%%%%%%%%%%%%

\subsection{The angular momentum content}
\label{SecAngularMom}
%The angular momentum content is a key to spiral galaxies aspects. The standard model for disc formation is based on the global angular momentum conservation. The scale relations  are naturally reproduced in this scenario.
As mentioned in Section \ref{sec:intro}, different mechanisms could simultaneously act to shape the $\Sigma(r)$ profiles. Here, we explore the connection between disc type and    degree of  galaxy rotation.
To analyse this, we use the D/T ratio \citep{tissera2019}, the stellar spin parameter, $\lambda$, and the ratio between the  rotational velocity, $v$, and the dispersion velocity, $\sigma$,  of  galaxies, $v/\sigma$ \citep{Lagos2018}. All these parameters can be used to quantify the degree of stellar
rotational support of galaxies. We highlight that, while D/T values are estimated using  3D information of the particle distributions and their angular momentum content, the $\lambda$ and  $v/\sigma$ are 2D parameters obtained from a mock catalogue constructed by \citet{Lagos2018}. From this catalogue, in this work we adopted the $\lambda$ and $\sigma$ values obtained when  taking edge-on orientations of the galactic discs. These parameters are  r-band luminosity-weighted  within half-mass radius.  
Hence,  $\lambda$ and $\sigma$ provides a more direct comparison with observations while D/T ratios measure the actual degree of rotation.

%The definition of the  $D/T$ ratio is also related with the angular momentum of the galaxies (see Section \ref{Simulated}). 
In Fig.~\ref{figlambda}, the D/T distribution for the three $\Sigma$ types are shown. As can be seen, TII discs tend to have larger contributions of galaxies  with $D/T > 0.70$ while TI and TIII have larger number of galaxies with $D/T <0.65$.  The median D/T for the three samples are: $0.64\pm 0.08$, $0.68 \pm 0.08$, $0.63 \pm 0.07$ for TI, TII, TIII, respectively (the errors correspond to the standard deviation). These trends are in agreement with observations that reported a trend for  TII discs to be more frequent in late-type galaxies while TIII profiles,  in more dispersion-dominated  disc galaxies \citep{erwin2008,potru2006,gutierrez2011,debattista2017}.
%In this figure we also show the stellar mass distribution for each disc type subgroup. As can be seen the stellar mass distributions are very similar assuring that the trend mentioned before is not affected by a biased in the stellar mass distributions. 

%\begin{figure}
%  \includegraphics[width=\columnwidth]{Images/Parametros/HistogramDTstars}
  %\includegraphics[width=0.45\textwidth]{Images/Parametros/HistogramMassTypes}
%\caption{Fraction of galaxies with TI, TII and TIII discs as a function  of  $D/T$  The dashed lines are the  median D/T for each type of $\Sigma$ profiles. %Lower panel: the Fractional distribution of stellar masses for each disc type (the dashed lines are median of stellar. Histograms are normalized by the total elements in each subsample.
%}
%\label{morphology}
%\end{figure}

Values for the spin parameter $\lambda$ and the $v/\sigma$ distributions of galaxies for each  disc type are taken from \citet{Lagos2018}. Clear disc-dominated systems are expected to have $\lambda > 0.6$ approximately. However, in the middle panel of Fig.~\ref{figlambda} we can see an small tail towards lower $\lambda$ values which can be explained by the fact that these two parameters are estimated along the line-of-sight of edge-on projections \citep[see also figure 4 in][]{Rosito2019}. In Appendix \ref{lambda} we discuss the relation determined by these two parameters in more detail.
The $v/\sigma$ provides an alternative quantification of the level of rotational support of galaxies. As can be seen from  the middle panel of Fig.~\ref{figlambda}, the three $\lambda$ distributions are skewed to low $\lambda$ values, with the median values at  $0.64\pm0.15$, $0.73\pm0.12$, $0.70\pm0.1$  for TI, TII and TIII,  respectively. The errors correspond to the standard deviations of each subsample. TII galaxies tend to  have larger $\lambda$ parameters while TI and  TIII profiles tend to have larger contributions of galaxies with smaller $\lambda$. 
%In the case of TIII, this  finding is agreement with the observational results reported by \citet{wang2018}, where type TIII galaxies are related to  low spin $\lambda$ of the inner stellar disc (and a more extended
%HI disc which is not analysed this paper). 
In the lower panel of Fig.~\ref{figlambda}, the distributions for $v/\sigma$ are shown  with similar results. The median values of the $v/\sigma$ distributions are  $0.85 \pm 0.3$, $1 \pm  0.32$, $0.95 \pm 0.24$ for TI, TII and TIII.

To assess the statistical significance of these findings, we applied a Kolmogorov-Smirnov test (KS test) to the $\lambda$, $v/\sigma$, D/T distributions. %to compare two samples and provide estimations on how different these two distributions are by means of a parameters called p-value, if the p-value lower than a significance level (generally $\sim 0.05$), we can reject the null hypothesis (the two samples are drawn from different distributions).
Table \ref{table_KS}  shows the p-values corresponding to the comparison between the different distributions shown Fig. \ref{figlambda}. It provides, for different parameters, an assessment of how different/similar the samples are between them. In the case of D/T, the KS analysis is consistent with the trend for TII to prefer disc-dominated galaxies. 
%{\bf Similar distributions are detected between TII and TIII for $\lambda$ and $v/\sigma$ parameters, where in both $\Sigma(r)$ types are more concentrated in high $\lambda$ values. However for TI and TIII have a similar distributions with D/T parameter, which both $\Sigma(r)$ types tend to low values of D/T.}

TI and TIII show a trend to prefer galaxies less rotationally supported and, likely, with more significant inner spheroidal components or bulges. Conversely, disc-dominated galaxies  have a larger fraction of TII profiles. We note that these are only global trends since there is a variety of galaxies with different morphologies for a given  disc type. This finding suggests that, although the angular momentum content of the galaxies is relevant to shape the stellar mass distribution, and its  disc scale-length, it is certainly not the only physical mechanisms at play.
%Statistical differences are detected between TII discs and both TI and III discs, while no signal is found between TI and TIII discs.
%for TI and TIII, the p values are large for lambda, and v/sigma and is very small for D/T. 

 \begin{table}
   \begin{center}
   \caption{Statistical analysis:  p-values for the KS test applied to the $\lambda$, $v/\sigma$, D/T and stellar mass distributions for each type discs:TI, TII and TIII.}
 \label{table_KS}
 \begin{tabular}{lrrrr}\hline
  Disc type & p-value$_{\lambda}$ & p-value$_{\rm {v/\sigma}}$ & p-value$_{D/T}$ \\

 \hline
 TI vs TII  & 0.0006 &  0.0003 & 0.01 \\
 TI vs TIII & 0 & 0 & 0.43 \\
 TII vs TIII  & 0.39 &  0.08 & 0.0003 \\

 \end{tabular}
  \end{center}
 \vspace{1mm}
 \end{table}
 
% TI vs TII  & 0.04 &  0.01 & 0.01 \\
% TI vs TIII & 0.376 & 0.365 & 0.43 \\
% TII vs TIII  & 0.039 &  0.016 & 0.0003 \\

%\textbf{Fig. \ref{figlambda}\\
%Lambda\\
%TI vs TII : Ks2sampResult(statistic=0.197, pvalue=0.04)\\
%TI vs TIII : Ks2sampResult(statistic=0.077, pvalue=0.376)\\
%TII vs TIII : Ks2sampResult(statistic=0.202, pvalue=0.0385)}\\

%\textbf{
%V/Sigma\\
%TI vs TII : Ks2sampResult(statistic=0.229, pvalue=0.01)\\
%TI vs TIII : Ks2sampResult(statistic=0.078, pvalue=0.365)\\
%TII vs TIII : Ks2sampResult(statistic=0.224, pvalue=0.016)}\\

\begin{figure}
  \includegraphics[trim={0.5cm 1.5cm 1.5cm 2cm},clip,width=\columnwidth]{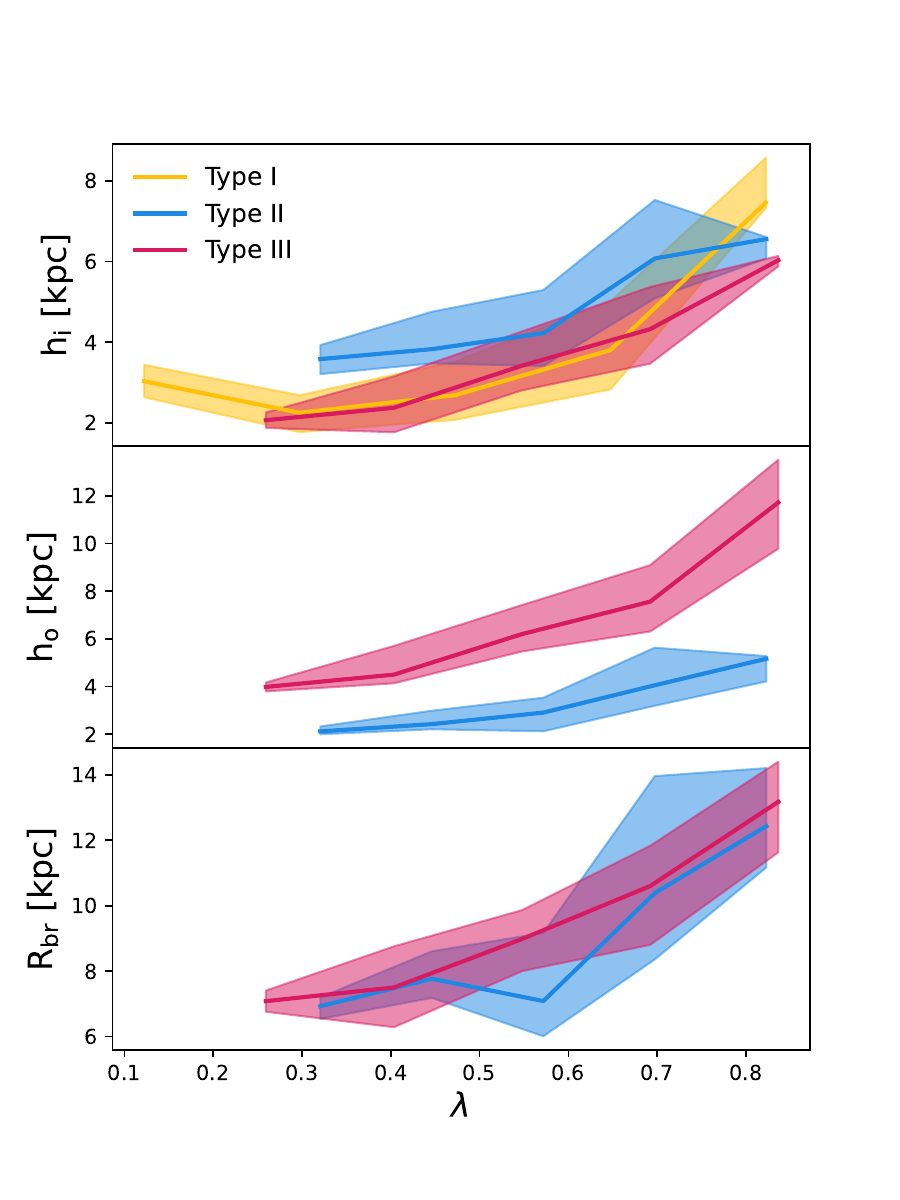}
  \caption{Median characteristic scales for the three disc types as a function of $\lambda$: $h_{\rm i}$ (upper panel), $h_{\rm o}$ (middle panel), and \rbreak (lower panel). The shaded regions denote the 25-75$^{\rm th}$ percentiles.}
  \label{scaleslambda}
\end{figure}

To get further insight into the relation between the degree of rotation and the disc types, in Fig.~\ref{scaleslambda} we show the inner scale-lengths, the
outer scale-lengths and the \rbreakk\;as a function of $\lambda$. We note that for TI discs only one scale-length can be defined and hence, there is no \rbreakk\;for these galaxies. Similar trends are found if $v/\sigma$ is used instead of $\lambda$.  From the upper panel of Fig.~\ref{scaleslambda}, it is clear that, galaxies with $\lambda \gtrsim 0.35$, $h_i$ increases with increasing $\lambda$, for all types of discs.
This is the expected trend when discs form under global angular momentum conservation \citep[][]{mo1998}. However, the trends change for $\lambda \lesssim 0.35$ in TI discs, where an anticorrelation is detected. We estimated that the Pearson correlation is $r = 0.55$ for $\lambda \gtrsim 0.35$ and $r =-0.35$ for $\lambda \lesssim 0.35$, for TI discs. Galaxies in this last range of $\lambda$ have  less significant discs than bulges and they might have been more affected by mergers and interactions \citep{Lagos2018,Rosito2019}.  
From this figure, we can also see that, at a given high $\lambda$, $h_i$ increases systematically from 
TI, TIII and TII discs. This suggests the action of different processes during  disc formation beyond  global angular momentum  conservation,  such as  redistribution of  gas and/or   stellar populations due to e.g. the accretion of satellites or disc perturbations like bars, spiral and lopsided modes.
The tendency for $h_i$ to be larger in TIII than in TI is agreement with the observational results reported by \citet{gutierrez2011}.

\begin{figure*}
    \includegraphics[width=\textwidth]{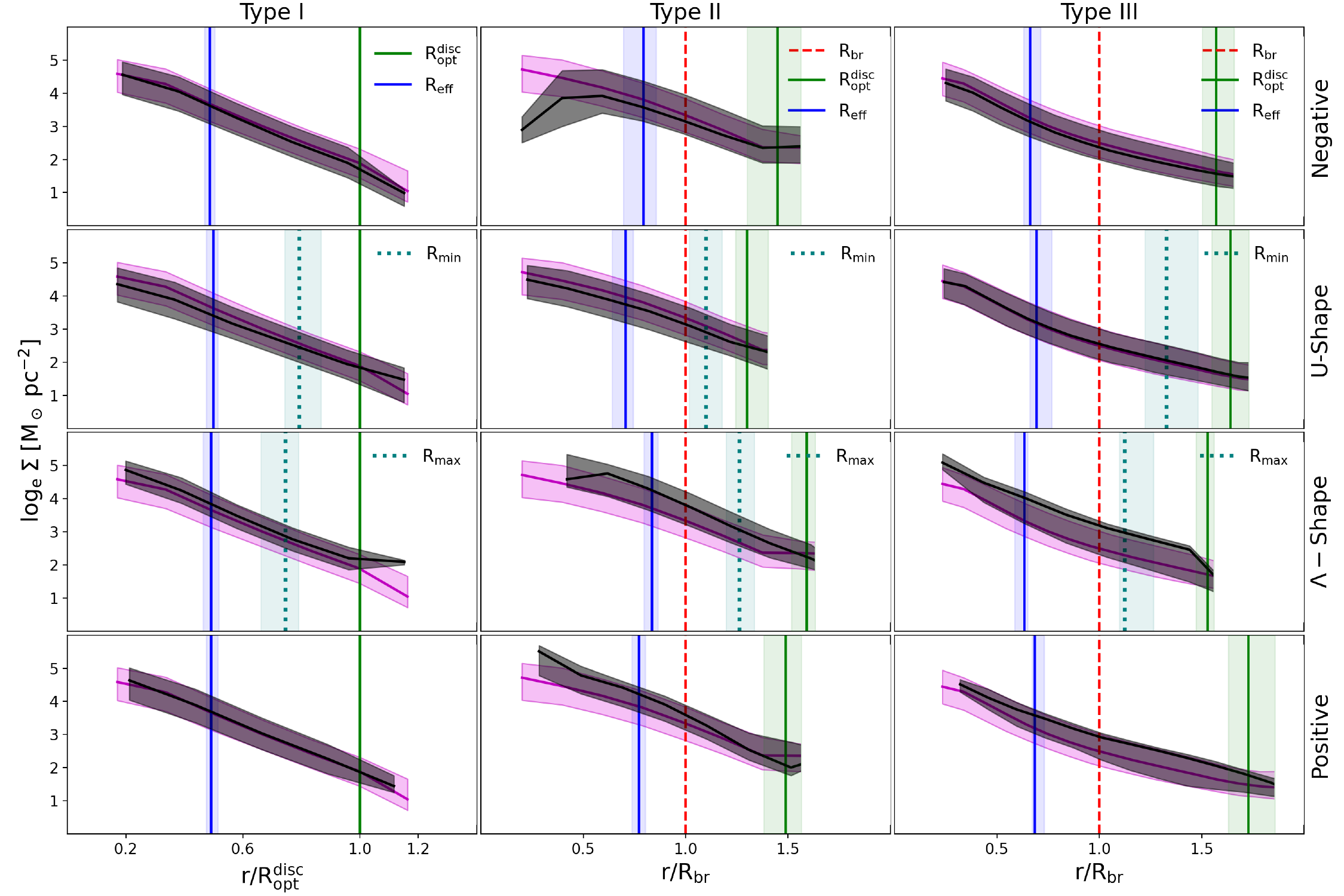}

  \caption{Median $\Sigma$ for TI (left column), TII (middle column) and TIII (right column) stacked profiles of discs with different age profiles (grey shaded region). For comparison, the median  $\Sigma$ stacked according only to  $\Sigma$ profiles are shown (violet shaded region). Shaded regions encompass the  $25-75^{\rm th}$ percentiles (and each profile staked were created between 0.5\reff to 1.5\ropt). The  medians and shaded regions defined by the $25-75^{\rm th}$ percentiles of \rbreak, \ropt, \reff, \rmin, \rmax are also depicted.} 
    \label{surface_mass}
\end{figure*}

Regarding TII and TIII discs,   we can extend the comparison to $h_o$ and $R_{\rm br}$ as shown in the middle and lower panels of Fig.~\ref{scaleslambda}.
As can be seen, the $h_o$ for TIII discs are systematically larger than those of TII discs at given $\lambda$. The $h_o$ of both disc types increase with increasing $\lambda$.
However, the processes that shape the $\Sigma$ profiles in the outskirts of strong TII produce
a different dependence of $h_o$ on $\lambda$.
TIII and TII discs show clearly correlations between $\lambda$ with $h_o$ and \rbreakk\;. We estimated   Pearson correlation factors of $r \sim  0.51$ for both relations for TIII discs  and for  TII discs, we obtained  $r = 0.48$ and $r = 0.44$, respectively.

The \rbreak values are similar for TII and TIII profiles at a given $\lambda$,
%, although TII discs show a weaker trend with $\lambda$. 
in both cases, showing a positive trend, so that more rotationally supported discs break at larger galactocentric radius.
%\textcolor{blue}{However, the relation between the outer scale-lengths and \rbreak will be different.}
%;is very increasing in the range $\lambda \gtrsim 0.6$. For TIII discs, \rbreakk\;increases with increasing $\lambda$}.

Our results show that TII discs are preferentially found in galaxies which are strongly dominated by rotation. 

%{\bf Both for $\lambda$ and D/T are parameters that give us different perspectives on the content of the angular momentum of a galaxy. The trends seen in this section show that the TI and TIII galaxies have a fairly similar level of significance to each other and a tendency to be slow disc dominance rotators. Which suggests that both types tend towards a similar environment. On the contrary, the TII galaxies do not show a level of significance with the parameters in the p-value of the KS test with any other type sigma, which indicates that their origin of their sigma profile is different from TI and TII. }
%\begin{figure}
%  \includegraphics[width=0.45\textwidth]{Images/Parametros/hi_vs}
%  \includegraphics[width=0.45\textwidth]{Images/Parametros/ho_vs}
%  \includegraphics[width=0.45\textwidth]{Images/Parametros/Rbr_vs}
%  %\includegraphics[width=0.45\textwidth]{Images/Parametros/L_ep}
  
%  \caption{Parameters of the exponential types in function of spin parameter $\lambda$.}
%\label{scaleslambda}
%\end{figure}

% %---------------------------------------------------------------
% %

\subsection{Statistical analysis of  $\Sigma$ and the age profiles}
\label{SecSigma}
As we mentioned above, three types of $\Sigma$  and four different types of age profiles are defined and identified in the EAGLE sample. 
In order to further explore the connection between  characteristics of the $\Sigma$  and the distribution of stellar age in the discs, in  
Fig.~\ref{surface_mass} we show  the median $\Sigma$  for a given disc type (violet lines): TI (left column), TII (middle column) and TIII (right column).  For each disc type subsample, we subdivided and stacked the $\Sigma$ profiles according to the corresponding four age profiles. The resulting median of each subpopulation is shown by a black solid line.
To perform the stacking of $\Sigma$ profiles, they were first normalised by a characteristic radius. For the double-exponential $\Sigma$, i.e. TII and TIII, the median distributions are normalized by \rbreakk\;, while for TI galaxies by \ropt. The vertical green and blue solid lines  show the medians of \ropt and \reff, respectively. \rbreak is shown with a dashed red line while both $R_{\rm min}$ and $R_{\rm max}$ ( depending on the $\Sigma$ and age type) with cyan dashed lines, respectively. These last two radii denote the location of the minimum and maximum of the age profiles, respectively. All shaded areas shown in this figure are defined by  the $25^{\rm th}$ and $75^{\rm th}$ percentiles of the corresponding distributions.

As can be seen from Fig.~\ref{surface_mass}, TI discs show  very similar $\Sigma$ for all age profiles. For the double exponential profiles (i.e. TII and TIII) the dominating age profiles are the negative and U-shape ones  (black solid lines).  The positive  and $\Lambda$-shape profiles show slightly higher densities than the global stacked profile. Certain trends can be also noticed in the distributions of the characteristic radii,  namely \rbreak, \ropt, \reff, \rmin and \rmax (the latter two only correspond to the U-shape and $\Lambda$-shape age profiles, respectively). For TI profiles, both \rmin and \rmax tend to be {located} within \ropt, whereas for TII and TIII these radii are located between \rbreak and \ropt. 
%In Fig.\ref{surface_mass} and in the following figures, we can see certain trends in the means of the characteristic radii such as \rbreak, \ropt, \reff, \rmin and \rmax. The latter two correspond to the U-shape and $\Lambda$-shape age profiles, respectively. We see that for TI profiles, both \rmin and \rmax tend to be within \ropt. For TII and TIII, these radii are located between \rbreak and \ropt. 

\begin{figure*}
    \includegraphics[width=\textwidth]{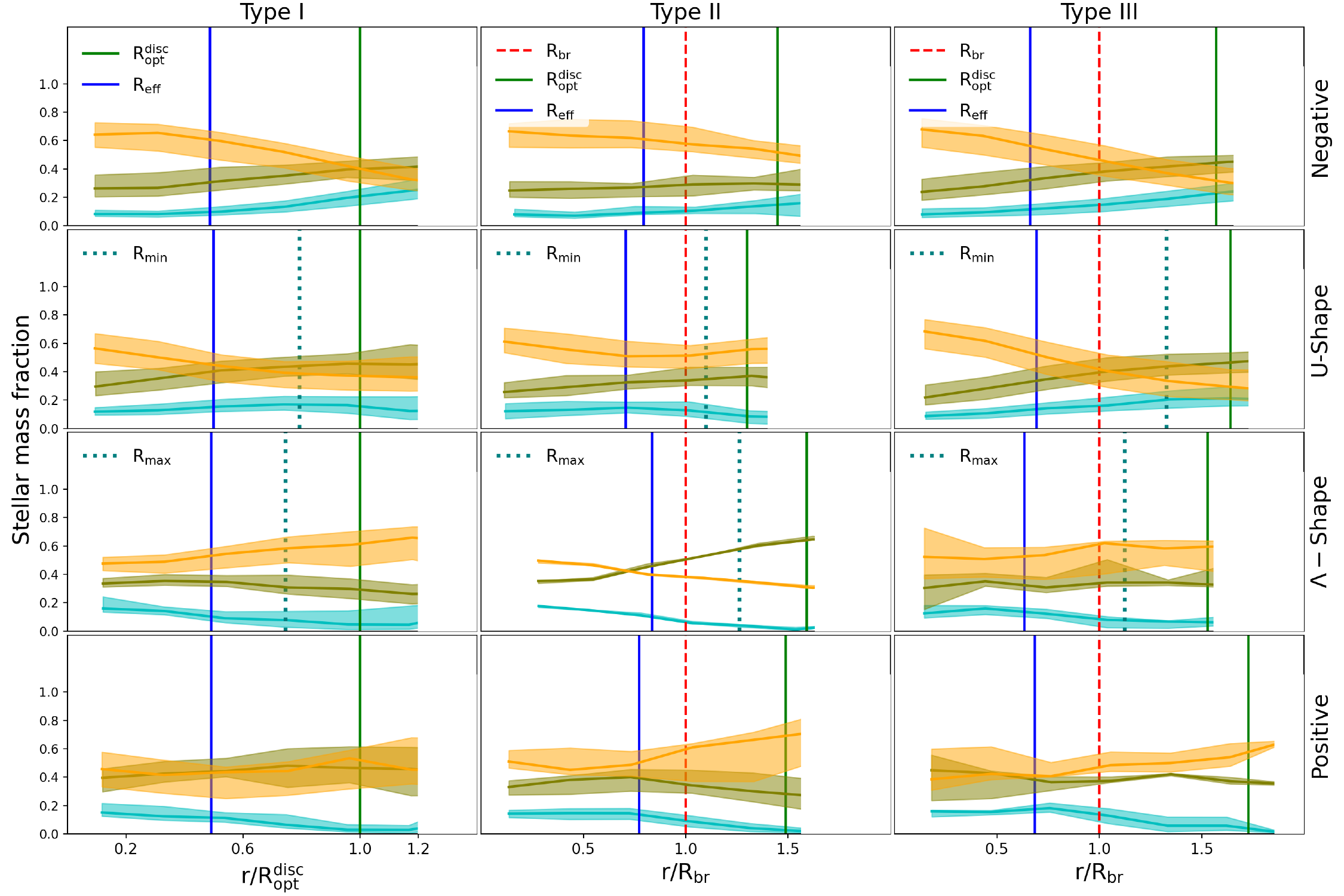}

  \caption{Median stellar mass fractions a function of the normalised radius for TI (left panels), TII (middle panels) and TIII (right
    panels) discs,  separated according to their age profiles: negative (upper panels), U-shape (second upper panels), $\Lambda$-shape (third panels) and positive (fourth panel).  
    Stellar populations are divided into three age groups:
    young stars, $<2$~Gyr (cyan lines), intermediate stars, $2-6$~Gyr (olive
    lines) and old stars, $>6$~Gyr (orange lines). The shaded regions are defined by the $25-75^{\rm th}$ percentiles. The following median characteristics radii are included: \rbreak, \ropt, \reff, \rmin, \rmax.} 
    \label{agetypes}
\end{figure*}

If we focus on the discs with the U-shape profiles, we note the upward bend of the age profile for TII galaxies (see $R_{\rm min})$ is located closer to  \rbreak than for the TIII type, where $R_{\rm min})$ tends to be at larger radii on average (but still within \ropt). In order to show how different the distributions are between these scalelengths, in Fig.~\ref{histos} we display $R_{\rm min}/R_{\rm br}$ distributions for TII and TIII discs. This suggests that TII discs with U-shape age profiles, where \rmin is closer to \rbreak in general,  experience a sharp decrease of star formation rate giving rise to an outer disc with older population. In the case of the TIII profiles the change of the star formation activity may occur further out in the discs. As a result the TIII discs could be populated by younger/intermediate age stars out to larger radii. In the following sections, we will analyse the distribution of stellar populations by age and the star formation rate density. We also acknowledge the fact that that other  processes such as mergers, radial migration or environmental effects could also contribute with old stars to the  outer regions of TIII, particularly.

\begin{figure*}
    \includegraphics[width=\textwidth]{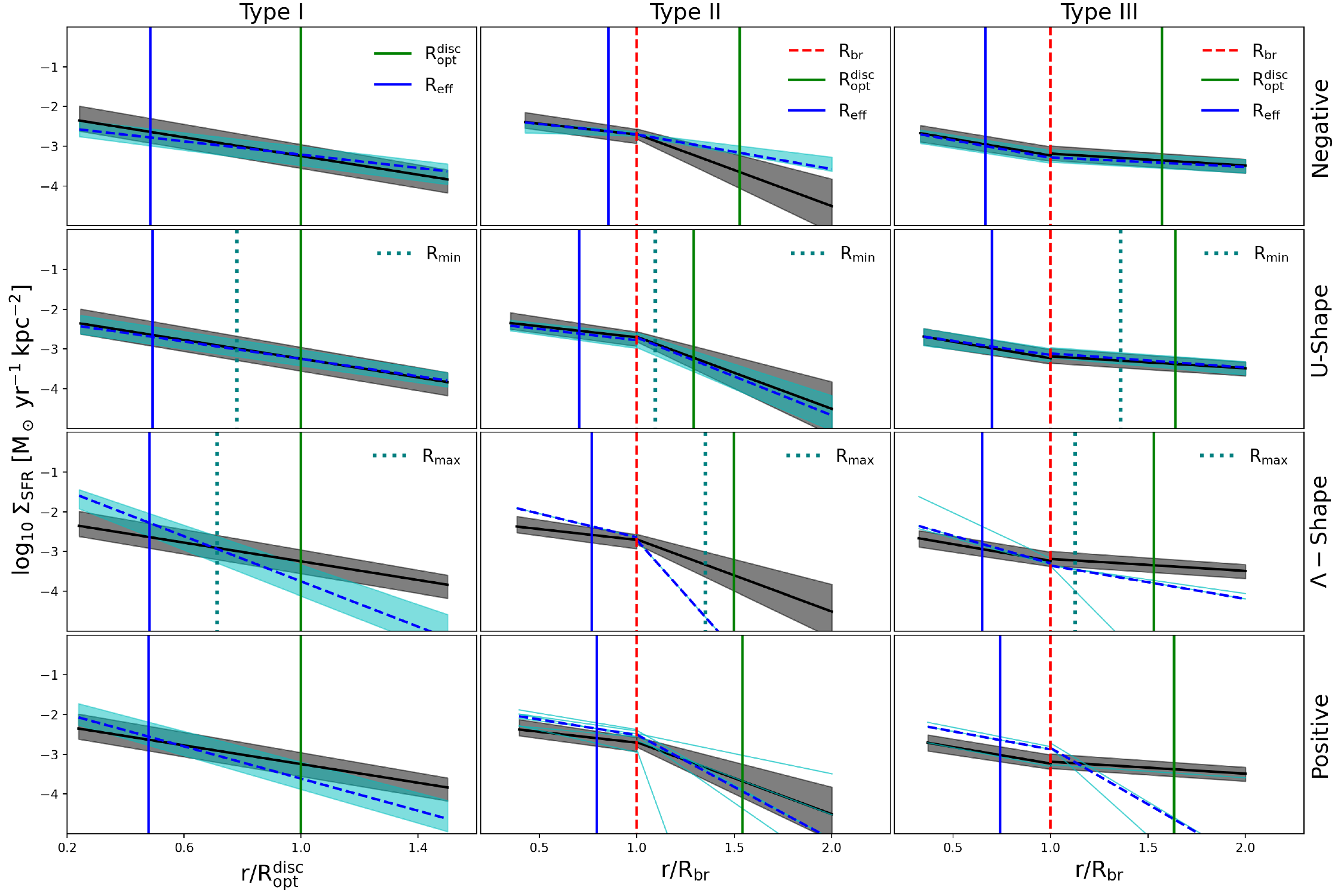}
      \caption{ {Median $\Sigma_{\rm SFR}$ profiles for different types of $\Sigma$ and  ages profiles (dashed blue lines). For comparison, the median $\Sigma_{\rm SFR}$ profiles of each disc type are included (gray solid line). The shaded regions cover 25-75$^{\rm th}$ percentiles of the corresponding relations (gray and cyan shaded areas)}. The characteristics radii,  \rbreak, \ropt, \reff, \rmin, and \rmax,  are  included when corresponding.}
  \label{SFRsurface}
  \end{figure*}

\subsection{Distribution of the mono-age stellar populations }

In order to understand the origin of the different $\Sigma$ profiles in
relation to the characteristics of their age distributions, we compute the stellar mass fraction profiles for each galaxy by considering three different age intervals: young stars ($<2$~Gyr), intermediate age stars (2-6~Gyr) and old stars ($> 6$~Gyr). They will be considered as three mono-age stellar populations. TI galaxies are normalised by \ropt, while TII and TIII galaxies are normalised by \rbreak. In this way, they can be stacked according to  galaxies  disc types and age profiles.
For each subsample, the median stellar mass fraction for mono-age stellar populations is estimated as a function of  radius. The mass fraction is calculated with respect to the total stellar mass per radial interval. Figure~\ref{agetypes} shows the stacked stellar mass fractions of discs for each of  $\Sigma$ type and each age type. The corresponding characteristics radii are also included. The shaded colored regions correspond to young, intermediate and old age intervals (cyan, olive and orange regions respectively), and are enclosed by  $25^{\rm th}$ and $75^{\rm th}$ percentiles of the stacked profiles each subsample. For these stacked profiles, \reff, \rbreak, \rmin and \rmax are the corresponding medians for  each $\Sigma$ and age types. Hereafter , we will only use these medians
for reference. %since in Section \ref{SecSigma} we made a discussion about the distribution of the scale radii of each $\Sigma$ and age types.

As can be seen from Fig.~\ref{agetypes},  TI profiles with the U-shape and negative age profiles tend to have  larger mass fractions of old populations than intermediate ones  in the central region. In fact, approximately 60 per cent of the stellar mass corresponds to old stars. The contribution of these mono-age populations decreases rapidly for increasing radius. Meanwhile, the fraction of young stellar populations varies slightly within \ropt, increasing systematically with increasing radius for discs with negative age profiles so that it reaches 20\% by \ropt.  For systems with  U-shape age  profiles, the  fraction of young stars are similar, with a weak increase around  \rmin. However,  at this characteristic radius  the contribution of intermediate age stars is more important than the old stars, explaining the change in slope of the age profile.
For $\Lambda$-shape and positive profiles, the relative fractions of the old and intermediate populations are dominant throughout the discs. However, they have  higher concentration of young stars in the central regions. This concentration diminishes {\rm significantly} at about \rmax for $\Lambda$-shape and close to \ropt for the positive profiles.

For TII discs with  U-shape and negative age profiles, the old population represents the largest mass contribution on the whole disc, representing more  60 per cent in the central regions. It is interesting to note the change of trends in the old populations from \rmin in U-shape profiles. Intermediate age stars have an increasing contribution from the inner to the outer regions in U-shape age profiles, while negative age profiles is almost constant. In particular, the mass fraction of young populations for negative profiles tend to be almost constant within \ropt. However, for systems with U-shape profiles, there is clear decrease of the contribution of young population from \rmin. The latter suggests a drop in star formation activity at large galactocentric distances. For galaxies with positive age profiles, there is an  important contribution of old stars populations in the outer region ($r >$\rbreak), and a decrease of young and intermediate populations. Conversely, the mass fractions of   young and intermediate stellar populations increases in the central regions. However, we stress that for $\Lambda$-shape age profiles the number of galaxies is very small to draw a robust conclusion.

For TIII discs  with negative and U-shape age profiles we also find larger fractions of old populations in the inner regions. The old disc decrease rapidly for $r >$\rbreak while the intermediate age populations have more contribution in this range. This might indicate the presence of an old inner disc which is more concentrated than in the TII. On the other hand, the outer discs are more dominated by intermediate age stars from \rmin  in U-shape and from \ropt  for negative age profiles. The young populations tend {to increase in the outer region, even are similar to the contribution of old stars in the very outskirt of the  discs}. Similar to TII, the low number of galaxies in the $\Lambda$-shapes and negatives profiles are not enough to draw a robust conclusion.

In summary, we find different contributions of mono-age stellar populations in the three defined disc types, which modulate their $\Sigma$ profiles.
We note the physical meaning of \rmin in U-shape age profiles, which might highlight the locus of a  drop in star formation activity in TII profiles (and probably the origin of the break in these types) and the dominance of the intermediate population, even over the old stellar population for $r >$\rbreak in  TIII galaxies. We will explore this issue in more detail in the following section.

\subsection{Statistical analysis of the star formation activity}

In this section, we explore the star formation rate surface density, $\Sigma_{\rm SFR}$, for galaxies with different $\Sigma$  and age profiles  to provide an interpretation on the origin of the behaviours described above.
To estimate the current star formation activity, the $\Sigma_{\rm SFR}$ is computed considering  stars particles younger than 2~Gyr. For this estimation, only  galaxies with more than 500 young stars particles were considered\footnote{{We acknowledge the fact that this is a  relative large age interval. However, adopting a shorter one results in significant reduction of the number of galaxies.  Consequently, it would affects our statistics. On the other hand, 2 Gyr is the same threshold assumed to define young stellar populations. This allow us to correlate the results of Fig. 6 and Fig. 7 more directly.}}. The $\Sigma_{SFR}$ profiles are estimated by using moving averages. The aim of this section is to analyse the changes of $\Sigma_{\rm SFR}$ profiles between the inner and outer zones of the discs.
For  TII and TIII discs, $\Sigma_{SFR}$ profiles are normalized by \rbreak while for TI discs, they are normalized by \ropt. Then we fitted a linear regression between $0.5$\reff and $1.5$\ropt for the TI galaxies, and applied a double linear regression to the TII and TIII galaxies (see Section \ref{BFAlg}), within the radial range [$0.5$\ropt- \rbreak] and  [\rbreak - $1.5$\ropt].  {We then estimate the median gradients and the  $25^{\rm th}$ and $75^{\rm th}$ percentiles for each corresponding subsample. }
The results of this procedure are shown in Fig. \ref{SFRsurface}. The black solid lines show the overall median $\Sigma_{\rm SFR}$ profile for  TI (left panels), TII (middle panels) and TIII (right panels) type discs, and has been included for reference. The blue, dotted lines show the median profiles obtained by  subdividing the samples according to their age profiles. The shaded regions (cyan) correspond to 25-75$^{th}$ percentiles of the corresponding subsample. Due to the low number of galaxies with enough young stars, TII and TII galaxies with positive and $\Lambda$-shape age profiles are shown individually.

%{\bf galaxies with enough young stars to fit in} TII and TIII galaxies with positive and $\Lambda$-shape age profiles, we show the individual profiles.} 

As can be seen from  Fig. \ref{SFRsurface}, the negative and U-shaped age profiles show a slightly lower $\Sigma_{\rm SFR}$ in the inner regions than the overall TI profile.
%with the negative profiles showing flatter trend the U-shape age profiles.
 This is agreement with our previous results that show that the old populations dominate the inner regions while  the intermediate stellar populations become more important  in the outer regions  as expected in a inside out formation model. Both types of age profiles are the most numerous in TI. Indeed, the general shape for TI galaxies (black region) is mainly determined by  the U- and negative age profiles. We note that the 
$\Sigma_{\rm SFR}$ profiles in TI discs with U-shape age profiles does not show a break,  in agreement with Fig. \ref{surface_mass} where we can see that the intermediate age stars seem to be responsible of the shape of age profiles. TI discs with positive or $\Lambda$-shape age profiles  show a clear change in the median slope of $\Sigma_{\rm SFR}$ with respect to the overall TI profiles, indicating  higher level of star formation activity in the inner regions than in the outer parts. TI galaxies with $\Lambda$-shape show the highest SFR density in the central regions and the lower one in the outskirts. 
 
 TII discs have mainly negative and U-shape profiles.  Those with negative slopes have higher $\Sigma_{\rm SFR}$ profiles than average for $r >$ \rbreak. These SFR profiles can be well fitted by a single exponential (the change in slopes between the $r >$ \rbreak and $r <$ \rbreak is very small). TII systems with U-shape age profiles have a SFR distribution consistent with $\Sigma$ profiles, i.e. they show a clear break at about the \rmin.  The higher star formation activity in the central region keeps reinforcing the break in the $\Sigma$ profiles.
  In TII discs there is only one $\Lambda$-shaped age profile, indicating that the rest of the galaxies in this group have few young stellar particles. These systems have higher SFR in the inner region and a sharp decrease for $r >$~\rbreak. We emphasize that the change of slopes of the age profiles at \rmax is due to higher contributions of the intermediate and old stellar populations in the outer regions. 
 Finally those discs with positive profiles have higher inner SFR and slightly lower one in the outer parts. 
 
 TIII discs show $\Sigma_{\rm SFR}$ profiles with similar breaks and change of slopes for discs with negative and U-shape age profiles. The change in slope is consistent with a slight increase of the SFR in the outer regions, for $r > $\rbreak, but the \rmin is further away from \rbreak. This could suggest that  U-shape profile might be modulated by other processes beyond the star formation activity in the galaxy. Those with positive and $\Lambda$-shape age distributions show again  higher SFR in the central regions as expected. We recall that there are fewer than 10 galaxies in these subsamples. Hence TIII tend to have negative and U-shape profiles, as mentioned in the previous section. The U-shape in age profiles can be associated to  both a decrease of the SF activity with decreasing radius and the more significant contribution of intermediate age stars in the outskirts as discussed in the previous section. 

\begin{figure*}
    \includegraphics[width=\textwidth]{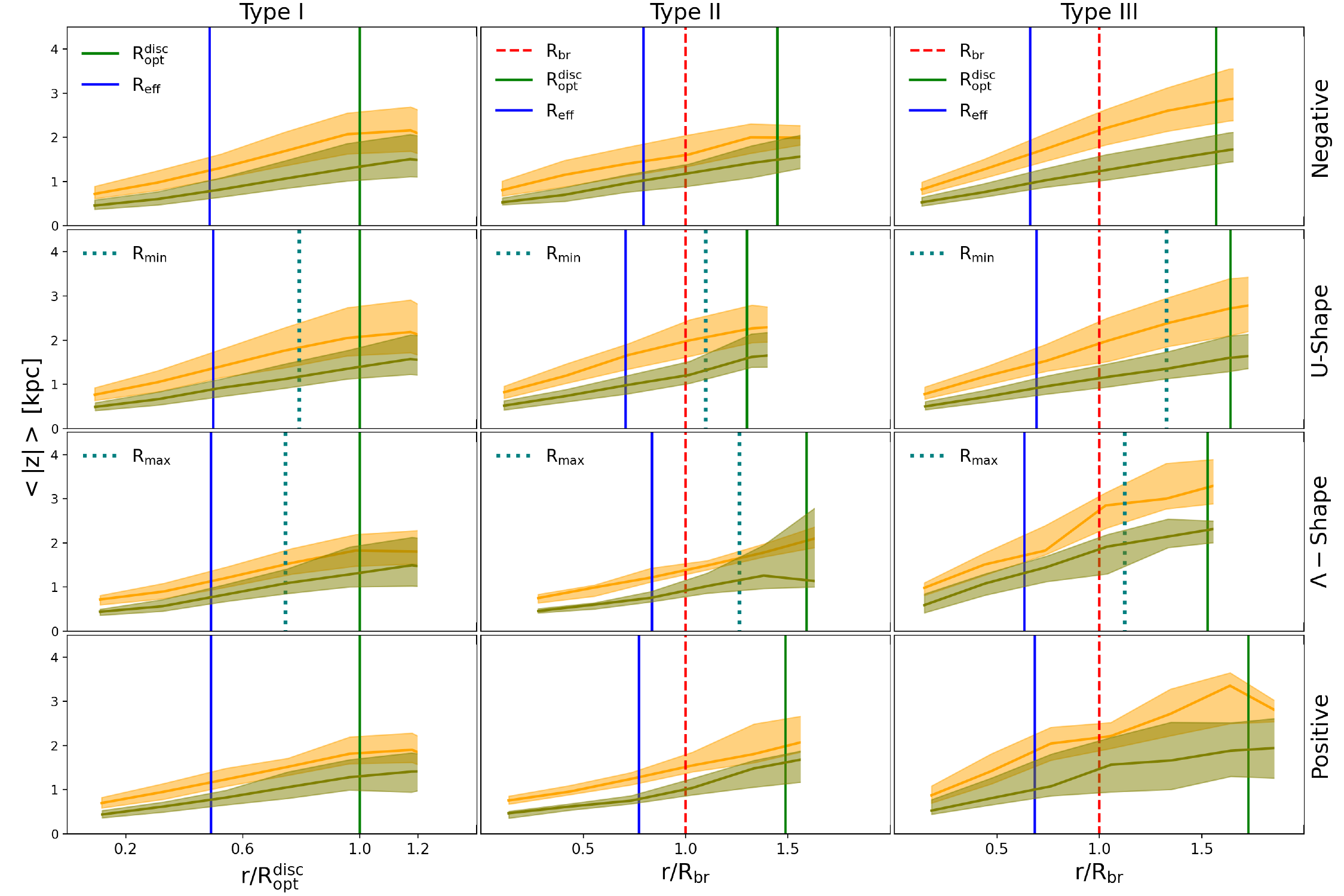}
     \caption{Distribution of the median height ($<|Z|>$) with respect to the midplane galactic as a function of galactocentric distances. {These profiles are stacked according to the $\Sigma$ and  age types. The shaded regions are enclosed by  corresponding 25-75$^{\rm th}$ percentiles. The orange shaded regions correspond to the old stars with ages $> 6$ Gyr and the olive shaded regions,to the intermediate age stars, 2-6 Gyr.} }
     \label{ZZPro}
\end{figure*}
%{\bf In general, the strong types sample show that there are a change of slopes in the $\Sigma_{\rm SFR}$ for double exponential galaxies, which outer TII disc tend to have the same or lower $\Sigma_{\rm SFR}$ than their inner region. TIII galaxies show a slightly increase of $\Sigma_{\rm SFR}$ in the outer zone in relation to the inner disc (this trend are not fulfilled in positive age profiles). The $\Sigma_{\rm SFR}$ for TI galaxies decrease exponentially with the galactocentric radius,  this suggest that have greater of $\Sigma_{\rm SFR}$ in inner regions than the outskirts of the discs. Due to normalization by \ropt is not possible to determine whether TI galaxies are transition between TII and TIII in those profiles.}

 {In summary,  the  $\Sigma_{\rm SFR}$ profiles vary differently for different disc types. TI discs have $\Sigma_{\rm SFR}$ gradients that vary for different age profiles}, being negative steeper for discs with positive gradients. TII and TIII discs have $\Sigma_{\rm SFR}$ profiles  that change the slopes towards steeper negatives  and shallower negative values for $r>$\rbreak, respectively. In the case of TIII profiles the break in the U-shape age profiles are further out, overall,  in comparison with the break in the $\Sigma$ profiles, suggesting that the other mechanisms could contribute to shape them.

\subsection{Statistical analysis of the median vertical height profiles}

 { Radial migration via scattering by non-asymmetric perturbations or churning might have an impact on the outskirt of disc galaxies \citep{debattista2017}. {The contribution of stellar migration to vertical heating in the outskirts and secular evolution in disc galaxies could also lead to flaring} \citep{minchevandfamaey2012}}, although mergers and interactions are also proposed to be the main cause for the effects \citep{garciadelacruz2021,Grand2017Au}. Disc flaring increases with radius in older stellar populations, while flaring in younger populations is reported to be weaker \citep{minchevandmartig2015}. {Thus, a thick disc component and a high velocity dispersion of the old stellar populations in the disc outskirts could indicate the action of scattering \citep{Minchev2010} or radial migration \citep{minchev2012}}. Nevertheless the effect of radial migration on discs is still a controversial issue with many works providing different results \citep{roskar2013,grand2016}. 
 %Additionally, churning could also induce radial migration, however, this mechanism is not expected to increase the vertical dispersion in the disc outskirts \citep{shellwoodbinney2002}. 

While it is beyond the scope of this paper to study in detail the impact of these mechanisms\footnote{We will not analyse in detail the displacement of stellar particles from their birth radius due to the low number of snapshots available to follow the orbits. This introduces numerical noise which does not allow a clear estimation of the fraction of migrated stars.}, we can assess the behaviour of the flaring in the median absolute heights, $<|z|>$, for different types of discs. For this purpose we estimate the heights of the old ($>6$ Gyr) and intermediate (2-6 Gyr) age stellar populations with respect to the galactic mid-plane as a function of radius (normalized by \rbreak in double exponential types and \ropt for TI discs) for the different disc types. Then, the distributions are stacked per $\Sigma(r)$ and age profile types. 

In Fig. \ref{ZZPro} we show the $\rm <|z|>$ profiles stacked for the old stars  (orange regions) and intermediate stars (olive regions). 
%The medians of <|Z|> of each age interval are estimated as a funtion of radius, the profiles of each galaxy are normalized by \ropt to correctly stack. This profiles also we separated between the $\Sigma$ types and age types.
%We show that those TIII galaxies with negative or U-shape in its age profile have a greater dispersion in its outskirt than type II galaxies. The dispersion in young populations are greater in TII galaxies than TIII. For all this comparation TI is a transition between TIII and TII.

For TIII discs the old stellar populations tend to have the larger flaring with respect to other two disc types. This trend is present for disc with different age  profiles. Discs with positive and $\Lambda$-shape age profiles have larger variations  due to the low number of members (i.e. less than 10 galaxies in the subsamples). TI and TII discs have similar level of flaring but TII systems show very  low variety (quantified by the 25-75$^{th}$ percentiles). These trend suggest that TIII would be more affected by radial migration or
by the accretion of satellites in the outer parts \citep{debattista2017}.
%TII discs tend to have smallest flaring respect to the other $\rm \Sigma$ types. 

%In relation to the age types, negative and U-shape have a similar flare as positive and $\Lambda$-shape. With the exception of intermediate stars, which for these age types there are very few members and must be analyzed carefully.
%TI discs show an intermediate trend between TII and TIII discs. This trend is reproduced for negative and U-shape age profiles, which correspond to ~63 percent of TI. For positive and $\Lambda$-shape galaxies we can see that the level of flaring are similar between intermediate and old stars in the outskirts.}

%\begin{figure*}
%   \includegraphics[width=0.33\textwidth]{Images/Globales/Z/MedianZ_Global_PTI}
%  \includegraphics[width=0.33\textwidth]{Images/Globales/Z/MedianZ_Global_PTII}
%  \includegraphics[width=0.33\textwidth]{Images/Globales/Z/MedianZ_Global_PTIII}
%  \label{Z}
%  \caption{All stars}
%\end{figure*}

\section{Discussion and Conclusions}
\label{conclusions}

In this paper we explored the shape of the surface density  profiles, $\Sigma$, of disc-dominated galaxies selected from the largest-volume simulation of the {\sc eagle} Project. Our sample of simulated galaxies was required  to have $D/T>0.5$ and their disc components were classified according to their $\Sigma(r)$ as types TI, TII and TIII. {To obtain reliable results, we focused our statistical  analysis on the so-called strong type, i.e., those galaxies identified to have  the strongest breaks on their density distribution. }
%We focused our statistical analysis  on the strong types subsamples in order to get more trustworthy results.
We also analysed the age profiles, the SFR surface densities, the mono-age stellar distributions and variation of the vertical height and its dispersion of galaxies belonging to the three defined disc type. 
Our main results  can be summarized as follows:
\begin{itemize} 

    \item The {\sc eagle}  discs show the three expected $\Sigma(r)$ distributions so that they can be classified as TI, TII and TIII discs. We find that, in this simulation and for mass-weighted $\Sigma(r)$, TIII and TI are the more frequent types in our EAGLE sample. We note that the simulated galaxies can be located in different kind 
    of environment.
    
     \item Overall, there is a trend between the morphology (i.e. quantified by $D/T$) and the disc  types. TII discs are more frequent in late-type spirals, while TI and TIII tend to be detected in more early-type spiral systems. These trends are in agreement with observations \citep{potru2006,gutierrez2011}. However, we note  that there is a large variety of morphologies for galaxies with different disc types, which suggests the  action of other physical mechanisms on the distributions of the stellar populations in the discs.

    \item  {Our analysis shows a clear correlation between  the inner disc scale-lengths  and  the stellar spin parameter, $\lambda$ %(i.e.  $\lambda$ is a 2D estimation using the stellar populations within the half-mass radius) 
    for all three  disc types with $\lambda \geq 0.35$. Additionally, at a given  $\lambda$, there is a systematic increase of the inner disc scale-lengths from TI to TIII and  TII, in that order. Indeed, TII discs have the larger inner  disc scale-lengths  in agreement with observations. Regarding the outer disc scale-lengths, both TII and TIII  discs show a  positive correlation with $\lambda$ and that the outer scale-lengths of TIII are  systematically larger. This suggests that the outer disc regions in the TII discs  might have grown  by  a weak star formation activity, maybe fed by smooth accretion (Fig. \ref{SFRsurface}), whereas in TIII discs other mechanisms associated with the local environment, such as galaxy interactions, could have affected the distribution of stellar populations \citep{younger2007,laine2014}} or stellar migration of intermediate age stars.

    \item Four types of  age profiles are identified in discs regardless of their $\Sigma$ type: negative, U-shape, positive and the so-called $\Lambda$-shape age profiles. The U-shape is the most frequent type followed by the negative one. Together they represent $\sim 70$ percent of the strong types subsamples. The positive and $\Lambda$-shape age profiles are found mainly in TI and TII galaxies, together accounting for 37 and 34 percent, respectively.
    %Their origin can be linked to the relative distributions of stellar populations of different ages, the spin parameters of the galaxies and the SFR distribution on the discs.
    { While U-shape, negative, and positive  age profiles  have been reported in  previous works \citep[e.g.][]{debattista2017}, we also identify $\Lambda$-shape profiles in the  few simulated discs. In this simulation, this age profile is produced by a significant increase of the star formation activity in the inner disc regions, together with  a slight increase of the intermediate populations (2-6 Gyr old) in the outer regions. However, we note that previous studies of dispersion-dominated galaxies in the {\sc eagle} simulations have also reported larger star formation activity in the central regions than expected, which might suggest the need for a more efficient AGN feedback, which could prevent late star formation activity \citep{rosito2019b,Rosito2019,lagos2020}. This could be also true for our sample which is composed by disc-dominated galaxies.

    %Hence, it is very important to explore in observations if $\Lambda$-shape age profiles are present in nature or if stronger SN/AGN feedback are needed to quench at least part of the star formation activity in the inner regions of these galaxies 
    }

    %   Both  positive and $\Lambda$-shape are determined by the excess of star formation in the central regions. Previous works suggest that interactions with nearby galaxies or even the accretion of small galaxies could trigger it \citep{mihos1995,perez2011,sillero2017}.  The existence of the intermediate age population which is crucial to define the breaks of the age profiles could be the results of radial migration, a more recent star formation activity which was abruptly quenched or the accretion of small satellites. It is beyond the scope of this paper to elucidate the contribution of each of these processes. }
     
   %\item For TII and TIII profiles we find that, in all cases, the change of slopes in the $\rm \Sigma(r)$  occurs at  radii smaller than radius where the age profiles change slopes. 
   %\end{itemize}
   %$\rm R_{\rm min}$ (or $\rm R_{\rm max}$), which corresponds to the 
   
    %Here, we summarise the main characteristics of the TI, TII and TII discs: 
    %of each disc type in relation to their properties are summarized as follows:   
     %\begin{itemize}
      
     \item {TI discs show  the distributions of mono-age stellar populations  consistent with an inside-out formation scenario, and have $\Sigma_{\rm SFR}$ profiles that decay systematically with increasing galactocentric distances. The U-shape profiles can be associated to larger contributions of intermediate age stellar populations. In fact, the minimum median age is statistically located at a radius from which the frequency of intermediate stars surpass those of the others mono-age stellar populations. TI discs with positive  and $\Lambda$-shape age profiles have both similar fractions of old and intermediate  stars  and more significant contributions of young stars in the central regions. The larger star formation activity in the central regions is accompanied by a shirking of the SFR in the outer regions. The overall median $\Sigma_{\rm SFR}$ is well approximately by a single exponential law but with a shorter typical scale-length than TI with negative and U-shape age profiles.
     The triggering of star formation activity in the inner regions could have been produced by  recent interactions with nearby galaxies that  destabilized the gaseous discs, transporting material to the centre. This would feed the star formation and the bulge component. This is consistent with the fact that TI discs tend to be found in early-type spirals. These events together with radial migration could also explain the increasing flaring  detected in TI discs.}

    %\item TI discs:  {\bf Overall, the flaring increases with the radius at a lower height than the TIII, however the uncertainties are greater than the TII. With respect to age profiles, the most frequent are the U-shape and negative profiles. 
   % For these are associated to higher fractions of old stars in the inner regions and a higher fractions of intermediate stars in the outer regions. Comparatively, U-shaped age profiles have large fractions of old stars up to \rmin and negative ones up to \ropt. }
    %The $\Sigma_{SFR}$ decline exponentially with radius and there are no breaks regardless of the shape of the age profiles.
    %{\bf Both $\Lambda$-shape and positive profiles, the fractions of intermediate and old stars are dominant and similar at all radius, while fraction of young stars is more important in the central region, this is in agreement with $\Sigma_{SFR}$. The latter could have been by a recent interaction with another galaxies which destabilized the gaseous disc feeding inner star formation. This is consistent with the fact that this (and TIII) are found in early-type spirals.}
    
    \item  {TII discs  tend to have mainly negative or U-shape profiles. The inner regions are dominated by old stars whose fractions decrease steadily with increasing radius.  Those with negative age profiles have smaller fractions of young stars than those with U-shape ones. In the latter case, the median \rbreak  and \rmin are very similar, indicating that the break in the density profiles is immediately followed by a the change of slope in age distributions.  This change is produced by a lower contribution of young stars as a results of the sharper truncation of the star formation activity. The few TII discs found to have positive/$\Lambda$-shape age profiles have larger star formation activity in the central region, which is strongly truncated in the outskirts. Additionally, they have larger contributions of young stars in the inner discs as expected.
    TII discs also show flaring in the heights but with the smallest dispersion.}

    %\item TII discs: 
    %{\bf These galaxies present the smallest median height with respect to other $\Sigma$ types. }
    %Scattering cannot be discarded and, in fact, it could one possible mechanism to explain the origin of the intermediate populations.
%    {\bf Those systems with U-shape and negative age profiles tend to have larger fraction of old stars in the central region, however this contribution is less than other $\rm \Sigma$ types, which it could be a characteristic of late-type. The dominance of intermediate stars in the outer zone is softened by old stars. Comparatively, $\Sigma_{\rm SFR}$ in  U-shape profiles decrease sharply in the outer disc, while for the negative profiles this is more flattened, which the star formation is more active. {\it Both $\Lambda$-shape shape and positive age profiles have  a significant contribution of intermediate and old stars at all radius. However the young star only show a high fraction in inner regions, this in agreement with $\Sigma_{\rm SFR}$, where these profiles show a slightly higher star formation in this region and lower $\Sigma_{\rm SFR}$ in the outskirts}}

    \item {The up-bending of TIII discs coincides with  a sharp decrease in the fraction of old stars that dominate within \rbreak. For $r>$~\rbreak intermediate and young stars contribute more significantly than old ones. In case of U-shape and negative age profiles the fractions of young stars are almost constant as a function of radius. Additionally, the $\Sigma_{SFR}$ for  $r>$~\rbreak is higher than expected from a simple extrapolation of the inner $\Sigma_{SFR}$. The \rmin of the U-shape age profiles are located further out in the discs in comparison with those of TII discs. We also note that the fact that there are very few positive age profiles in TIII discs suggests that these discs could be more stable to perturbations that drive gas inflows to the inner regions and trigger central star formation activity,  reinforcing the speculation that accreted material from interactions or minor mergers in the outer regions could be a crucial mechanisms to shape the $\Sigma$ profiles (see also Fig.\ref{agetypes} where the contribution from different mono-age stellar populations are displayed).}

    %\item TIII discs: {\bf The old populations of these galaxies tend to have the largest median height with respect to other $\rm \Sigma$ types, regardless age profiles. In case of U-shape and negative age profiles tend to have largest fraction of old stars in inner region. On the outer region the intermediate stars are dominant, even more than the other $\rm \Sigma types$. The outer gradient of $\Sigma_{\rm SFR}$ tend to be slightly greater than inner gradient.}  Finally we note that there are almost no positive or $\Lambda$-shape age profiles. {\bf In the case of $\Lambda$-shape shape and positive age profiles, intermediate and old stars are dominant at all radius, however the younger population are more concentrated in the central region.}
    
    \end{itemize}

    %{ The  fact that the 
    %scale-lengths of TI discs correlate with $\lambda$ suggests that these discs could have formed by material accreted with global angular momentum conservation. The outer intermediate age stellar population could be part of a more recent  accretion (less than 6 Gyrs ago) which could have brought gas into the inner regions feeding a later star formation activity.} .
  
%\end{itemize}
%Our results agree with claims that the outskirts of galaxies store information that can be of large interest for the interpretation of forthcoming observations of low surface brightness systems (such as the LSST survey of the Vera Rubin Telescope) .
We  identified discs with the three expected $\Sigma$ types in the {\sc eagle} simulations, which can be associated with different  relative contributions of mono-age stellar populations, SFR activity across the discs and angular momentum content of the galaxies. In a future work, we will  focus on the role of environment, mergers and interactions.
%Our numerical statitical results will contribute to provide physical context for forthcoming observations of resolved stellar populations from JWST, which will improve the estimations of age and metallicities in the outer regions of disks.

\section*{Acknowledgements}
We thank the referee for useful comments which contribute to improve this paper.  SVL acknowledges financial support from the ULS through the "Beca Doctoral". FAG and SVL acknowledges financial support from CONICYT through the project FONDECYT Regular Nr. 1211370. FAG acknowledge funding from the Max Planck Society through a Partner Group grant. PBT acknowledges partial support from Fondecyt 1200703/2020 (ANID) and  ANID BASAL projects ACE210002 and FB210003.
This project was partially supported  by the European Union Horizon 2020
Research and Innovation Programme under the Marie Sklodowska-Curie 
grant agreement No 734374 and the ANID BASAL project FB210003. LAB acknowledges   support   from   CONICYT   FONDECYT/POSTDOCTORADO/3180359.
CDPL has received funding from the ARC Centre of Excellence for All Sky Astrophysics in 3 Dimensions (ASTRO 3D), through project number CE170100013.
This work used the DiRAC Data Centric system at Durham University, operated by the Institute for Computational Cosmology on behalf of the STFC DiRAC HPC Facility (www.dirac.ac.uk). This equipment was funded by BIS National E-infrastructure capital grant ST/K00042X/1, STFC capital grants ST/H008519/1 and ST/K00087X/1, STFC DiRAC Operations grant ST/K003267/1 and Durham University. DiRAC is part of the National E-Infrastructure. We acknowledge PRACE for awarding us access to the Curie machine based in France at TGCC, CEA, Bruyeres-le-Chatel.

\section*{Data availability}
The EAGLE simulations are publicly available; see \citet{mcalpine2016}



\bibliographystyle{mnras}
\bibliography{biblioSilvio.bib} % if your bibtex file is called example.bib

%\begin{thebibliography}{99}
%\end{thebibliography}

%%%%%%%%%%%%%%%%%%%%%%%%%%%%%%%%%%%%%%%%%%%%%%%%%%

%%%%%%%%%%%%%%%%% APPENDICES %%%%%%%%%%%%%%%%%%%%%

\appendix
\section{Distributions of \rmin, \rmax and \rbreak}

In order to better quantify the relative position of \rmin and \rmax of the age profiles with respect to \rbreak of the $\Sigma$ profiles, whose median values have been included in Fig. \ref{surface_mass}, \ref{agetypes} and \ref{SFRsurface}, we estimated the ratios \rmin~/\rbreak and \rmax/\rbreak for TII and TIII discs.

\begin{figure}
\begin{center}
 \includegraphics[width=0.95\columnwidth]{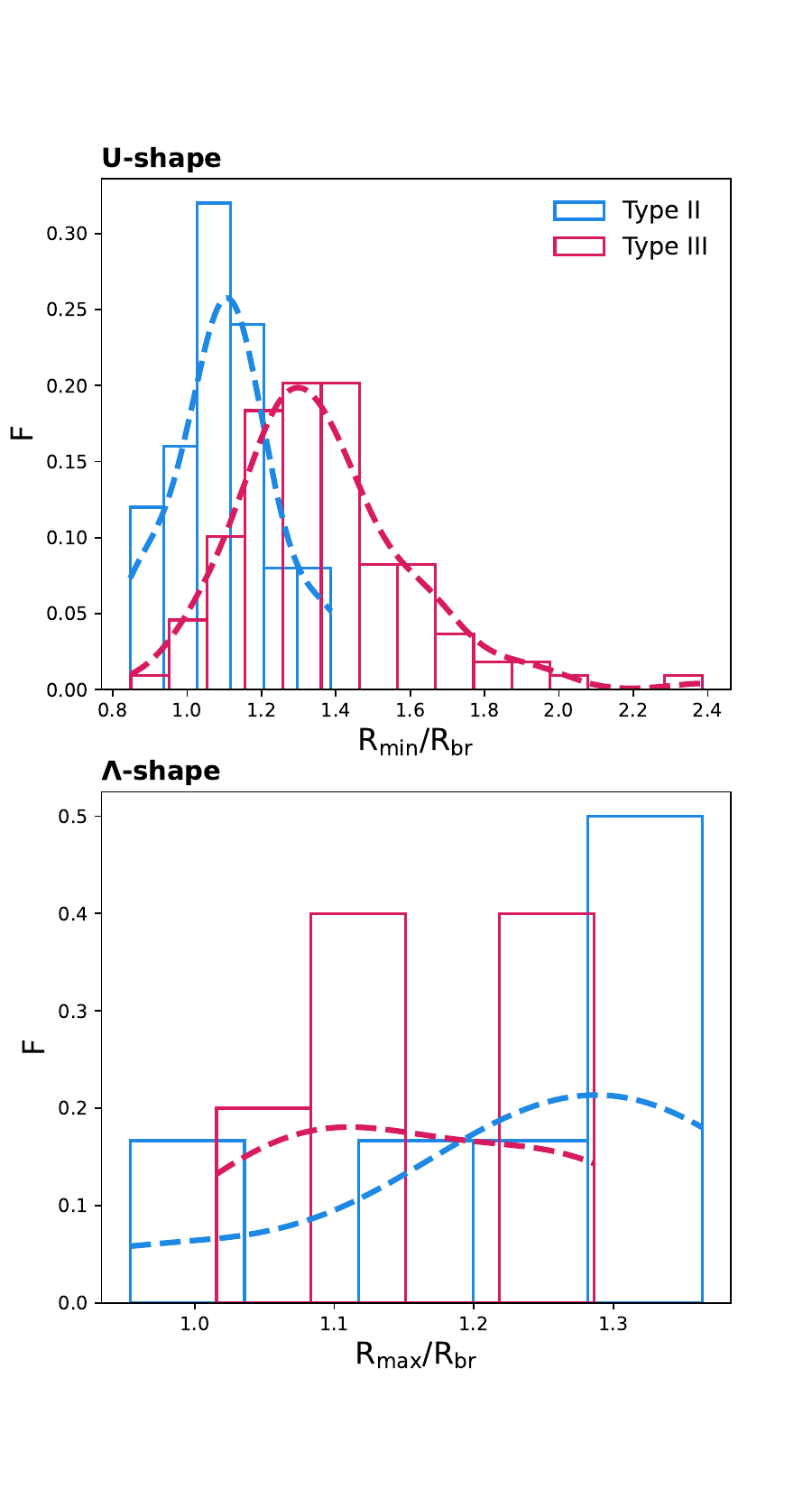}
 \caption{Distributions of \rmin/\rbreak (top panel) and \rmax/\rbreak (bottom panel) for U-shape and $\Lambda-$shape age profile, respectively, in  TII (cyan, dashed lines) and TIII (pink, dashed, lines) discs.}
 \label{histos}
 \end{center}
\end{figure}

In Fig.~\ref{histos} we show the probability density function (PDF) by using the Kernel Density Estimation. As can be seen from the top panel, galaxies with U-shape age profiles have different values and distributions of  \rmin~/\rbreak. TII discs show this ratio with a maximum frequency at  \rmin~/\rbreak$\approx 1$ and the distribution  are more concentrated around this value. Conversely, TIII discs have a more extended distribution, which cover a larger values of \rmin~/\rbreak,  indicating that in these discs the minima in the age profiles are located further out in the discs with respect to \rbreak. This justifies the conclusion that the upward bend of the U-shape age profile for TII galaxies is located closer to \rbreak than for TIII as shown in Fig. \ref{surface_mass}, \ref{agetypes} and \ref{SFRsurface} where the median values are displayed.

The larger variety of \rmin~/\rbreak detected for TIII discs,  most of them larger than 1, suggest that the physical mechanisms behind the change in age might not be directly related with the star formation process and hence, could be more associated to environmental effects or migration.

Similar estimations were performed for \rmax/\rbreak. As can be seen from the right panel of Fig.~\ref{histos}, the distributions are very noisy because of the low number of galaxies with this particular age profile. There is a weak trend for TII to have slightly larger values. However considering the low number of members of this subsample, we prefer to be caution and not draw a conclusion as discussed in the main text.

\section{Relation between  $\lambda$ and D/T}
\label{lambda}
We have adopted to main parameters to quantify the angular momentum of the galaxies: the  disc-to-total stellar mass ratio  and the spin parameter $\lambda$. The first one is estimated by using the full 3D distribution following \citet{tissera2012} as described in Section \ref{subsimulated_galaxy_sample}. 
%Hence, for each galaxy the stellar  discs
%are defined by star particles with $\epsilon= J_{\rm z}/J_{\rm z,max}(E) > 0.5$,  where  $J_{\rm z,max}(E)$ is the maximum angular momentum along the
%main axis of rotation, $J_{\rm z}$, over all particles at a given binding energy, $E$.
%In this paper we are using the data set of EAGLE galaxies selected by  \citet{tissera2019} where this decomposition method is applied.

The $\lambda$ spin parameter is taken from \citet{Lagos2018} and correspond to the r-band luminosity-weighted estimations along the line-of-sight towards galaxies oriented edge-on. These $\lambda$ provides a quantification of the angular momentum more comparable to observations.

\begin{figure}
\begin{center}
 \includegraphics[width=\columnwidth]{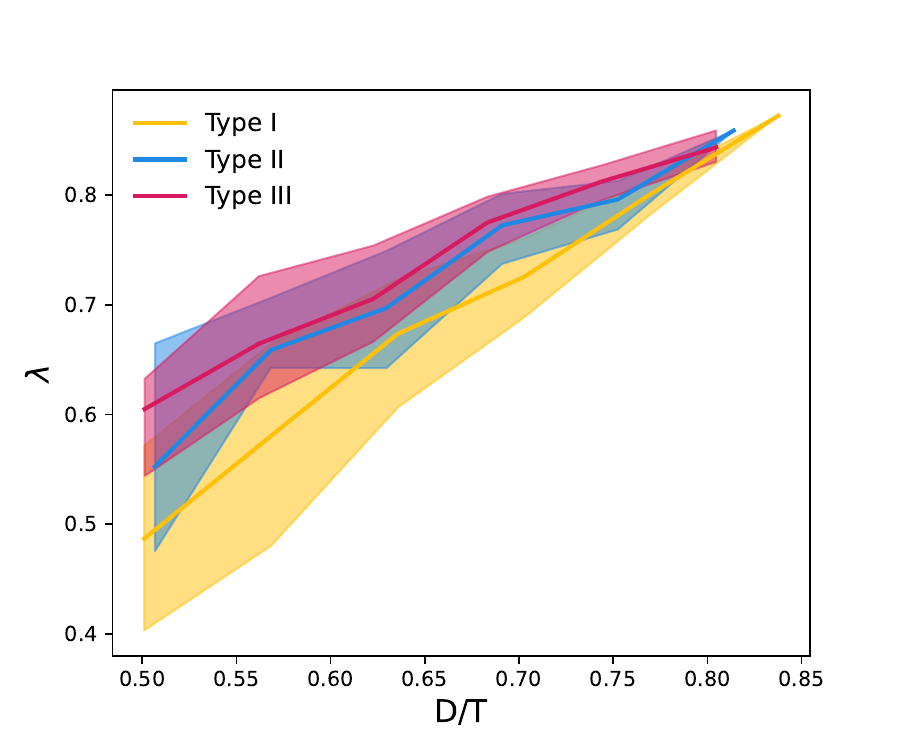}
 \caption{Correlation between $\lambda$ and $D/T$ for galaxies with TI (yellow), TII (blue) and TIII (pink) discs. The medians are shown by solid lines and the 25-75$\rm th$ are used to determine the shade areas.}
 \label{dtlambda}
 \end{center}
\end{figure}

Both parameters are used to show that the correlations between disc types and morphologies have physical meaning and the observed-motivated parameters show results comparable to observations. In Fig.~\ref{dtlambda} we show the relations between the these parameters for the three types of discs. As can be seen they trace each other very well with Pearson correlation factors of 0.65, 0.611, 0.67 for TI, TII and TIII discs, respectively.
We note, however, that the dispersion increases for decreasing D/T, i.e. as the bulge component becomes more important to equal the disc. In this range of D/T$ \approx 0.5$ there is more variation of $\lambda$ which could be due to projection effects. 

\label{lastpage}
\end{document}